# Pinning-mediated coalescence-induced lateral droplet motion on a nanotextured superhydrophobic surface

**Raushan Kumar, Gopal Chandra Pal, Chander Shekhar Sharma***

Thermofluidics Research Laboratory, Department of Mechanical Engineering, IIT Ropar, Punjab-140001, India

*Email: chander.sharma@iitrpr.ac.in, Ph: +91-1881-232358

**ABSTRACT**

Droplets coalescing on a superhydrophobic surface exhibit coalescence-induced droplet jumping. However, water vapor condensing on a superhydrophobic surface can result in the simultaneous formation of condensate droplets with two distinct wetting states, the Cassie state (CS) and partially wetting (PW) state. Droplets in PW state exhibit a high contact angle but are connected to the substrate through a thin liquid condensate column. Coalescence between CS and PW droplets has been recently identified as a possible mechanism for generating droplets exhibiting in-plane roaming motion during dropwise condensation on nanotextured superhydrophobic surfaces. Here, we systematically investigate this phenomenon through experiments on coalescence between sessile droplets in CS and PW states on a nanostructured superhydrophobic surface endowed with a micro-scale hydrophilic spot. Here, a sessile droplet carefully placed on the hydrophilic spot simulates the PW state. Overall, our investigations demonstrate that when a CS droplet coalesces with a PW droplet pinned to a hydrophilic defect, the interaction generates substantial in-plane momentum. We find that when the coalescing CS and PW droplets are nearly of the same size and about ~3 to ~3.5 times the size of the hydrophilic spot pinning the PW droplet, the vertical momentum generation is nearly completely suppressed, and the resulting maxima in in-plane momentum result in detachment of the merged droplet from the hydrophilic spot and its subsequent in-plane motion.

## 1. INTRODUCTION

Water droplet coalescing on pristine rigid superhydrophobic surface jumps vertically and passively sheds micro- and nanoscale droplets, enhancing applications such as self-cleaning,[1,2] condensation heat transfer performance,[3] anti-bacterial properties,[4,5] water desalination,[6] and anti-icing.[7] The motion of coalescing droplets on superhydrophobic surfaces exhibits intriguing variations depending on the scale of the surface structures.

On surfaces with microstructures, droplets do not always move vertically; instead, their trajectories can be oblique or shift laterally. This behavior is thought to arise from the interaction between droplets and the microstructure cavities, where coalescence along the side walls directs droplets in random, unpredictable paths.[8–12] In contrast, the lateral motion of coalescing condensate droplets is also observed on nano-textured superhydrophobic surfaces, with significant enhancement of heat transfer performance.[3,13] Experimental observations on dropwise condensation reveal that water vapor condensing on nanostructured surfaces inherently possesses wettability asymmetries due to the coexistence of the Cassie state (CS) and partially wetting (PW) droplets across the surface.[14,15] This phenomenon arises from the stochastic nature of nucleation within nanoscale surface features, where random variations in local topography and energy barriers lead to mixed wetting states.[3] It has been proposed that when droplets with different wetting states coalesce, the resulting adhesion asymmetry converts excess surface energy into lateral kinetic energy, driving lateral motion during the merging process.[3,6,13] However, a detailed experimental study is required to visualize the morphologies and hydrodynamics, and to quantify the generation of lateral motion as a result of droplet coalescence on nano-structured superhydrophobic surfaces.

In this work, high-speed video imaging combined with numerical methods has been employed to investigate the mechanism of the coalescence process for sessile droplets in different wetting states. A nanostructured superhydrophobic surface with a localized hydrophilic spot is fabricated. On this surface, a sessile droplet is placed on the superhydrophobic region to represent a Cassie-state (CS) droplet; meanwhile, another droplet is positioned on the hydrophilic defect to simulate a droplet in a partially-wetting (PW) state. First, the experiments are performed for the symmetric droplet coalescence, i.e., coalescence between droplets of the same size. Subsequently, the effect of asymmetry in the droplet coalescence dynamics is investigated.

## METHODOLOGY

### A. Preparation of a superhydrophobic surface with a hydrophilic spot

A glass slide is used as a base surface. The surface is subjected to a thorough cleaning process involving ultrasonication in acetone and isopropanol (SigmaAldrich), followed by a final rinse in de-ionized (DI) water. To enhance surface hydrophilicity and improve the adhesion and uniformity of the subsequent superhydrophobic coating, the sample is further treated with oxygen plasma (Harrick Plasma). Following this treatment, one face of the cleaned glass slide sample is super-hydrophobised by spray coating of commercially available solution of silica nanoparticles, homogenously dispersed in isopropanol solution (Glacco Mirror coat "Zero". Soft99.)[16]. Then, the surface wettability is characterized by measuring the contact angle on the coated surface using a goniometer (Holmarc Opto-Mechatronics Ltd, Model No- HO-IAD-CAM-01A-01A, India). The apparent advancing contact angle ($\theta_{ACA}$) and contact angle hysteresis (CAH) are obtained as $163° \pm 1.8°$ and $3.4° \pm 1.2°$, respectively. Subsequently, a hydrophilic defect is created by drilling a shallow blind hole of size $\sim 200$ μm and depth $\sim 10$ μm with the use of a microdrill tool.

**B. Experimental setup and procedure**

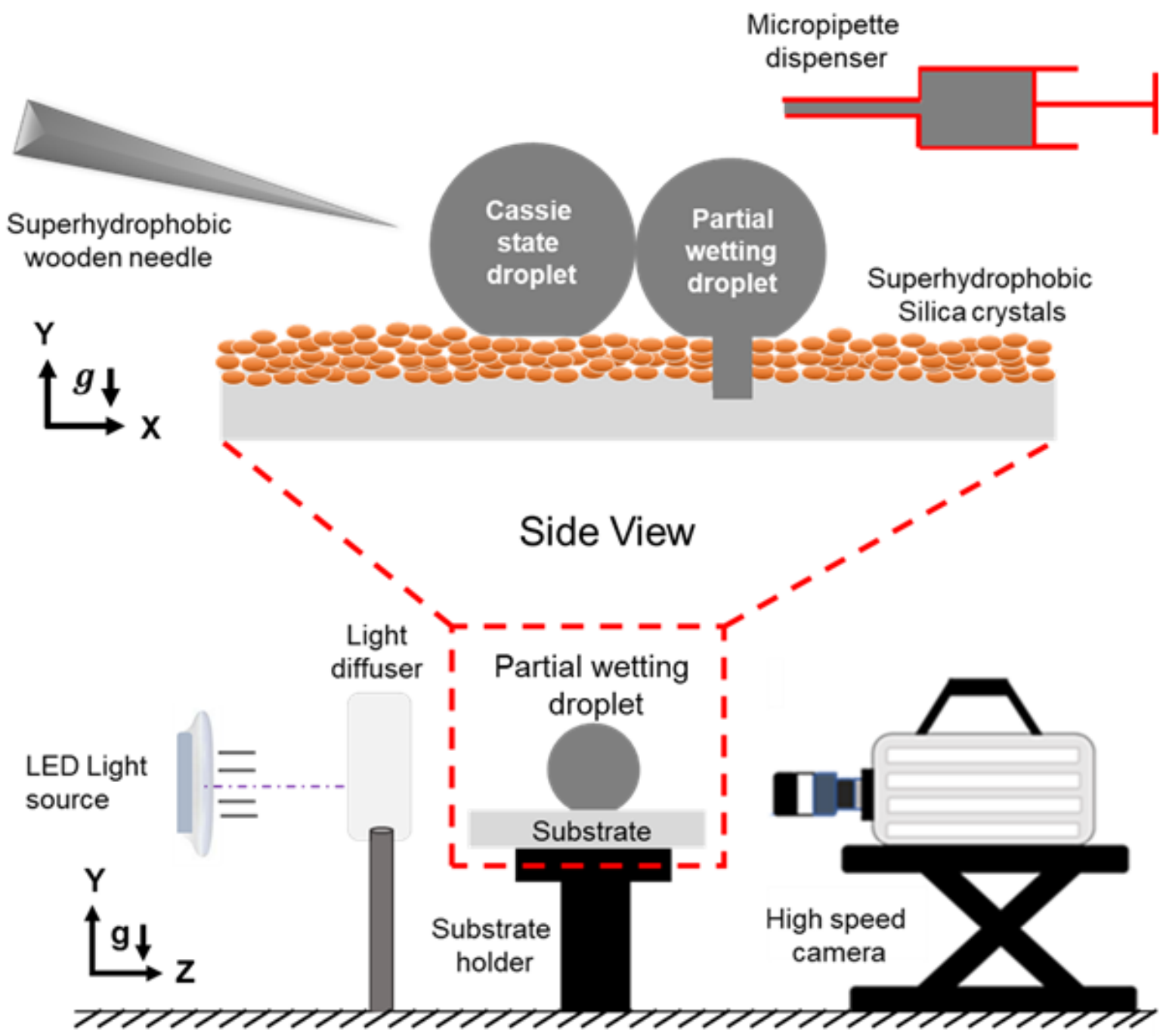

**Figure 1.** Experimental setup for Cassie state (CS) and partially wetting (PW) state droplets coalescence.

**Figure 1** shows a schematic of the experimental setup used for droplet coalescence experiments. The surface is securely mounted onto a surface holder using double-sided tape. To dispense the droplets on the surface, we used in-house developed superhydrophobic micropipette tips. These tips are fabricated through multiple dip-coating cycles in a nanoparticle-polymer solution, followed by heating at 80°C. The polymer solution consisted of 17 ml of acetone (Sigma Aldrich), 3 ml of Capstone ST 200 (DuPont Fluoropolymer Solutions), and 1gm of fumed hydrophobic silica nanoparticles (Evonik).[17] In each experiment, one microliter-sized DI water droplet (Cassie droplet) is carefully dispensed on the surface using superhydrophobic micropipette tips, and another drop (Partially wetting droplet) is carefully placed on the hydrophilic spot. An LED light source (AmScope) is used to illuminate the droplets and the coalescence process is captured through the high-speed camera (Fastcam SA4). To initiate coalescence, the dispensed Cassie droplet is gently moved into contact with the partial wetting droplet using a superhydrophobic wooden needle. The wooden needle is superhydrophobied by immersing it into an isopropanol solution (Glacco Mirror coat "Zero". Soft99.)[16] for 10 to 15 minutes. Image analysis confirmed that the kinetic energy imparted during this process is minimal, accounting for less than 0.01% of the droplet's surface energy.[18] The entire process is recorded using a Photron Fastcam SA4 camera at different frame rates ranging from 36,00 to 20,000 frames per second (fps) depending upon the size of droplets undergoing the coalescence process. Post-experiment, the ImageJ software is used to analyze the droplet movements from the recorded images.

**RESULTS AND DISCUSSION**

The droplet coalescence experiment is performed on the superhydrophobic surface with a hydrophilic defect to study the coalescence mechanism and measure the speed and direction of jumping droplets. Droplet coalescence experiments are conducted by carefully dispensing two microliter-sized water droplets onto the surface using a superhydrophobic micropipette tip. A droplet placed on the superhydrophobic region of the sample adopted a Cassie state (CS), characterized by minimal solid–liquid contact and the presence of air pockets trapped beneath the droplet. In contrast, a droplet deposited on a hydrophilic spot, microfabricated within a globally superhydrophobic surface, transitioned into a partial wetting (PW) state, exhibiting increased solid–liquid interaction. This enhanced interaction results in the PW droplet becoming locally pinned at the hydrophilic site, thereby restricting its mobility. To initiate coalescence between the two droplets, the CS droplet is carefully maneuvered into contact with the PW droplet.

**Figures 2(a), (b)** compare the coalescence of a droplet in CS state with another droplet in PW state and the coalescence of two droplets in CS state. The droplets are of the same size with a diameter ($D_0$) ~ 700 μm. The PW droplet experiences a localized pinning (adhesion) force, which effectively anchors it at the hydrophilic spot, making it act as a pivot point. This pinning force introduces an imbalance in the system, causing the PW droplet to encounter resistance in its motion compared to the CS droplet. Consequently, the coalescence process becomes asymmetric. In contrast, coalescence between CS-CS droplets proceeds symmetrically due to the absence of any adhesion force.

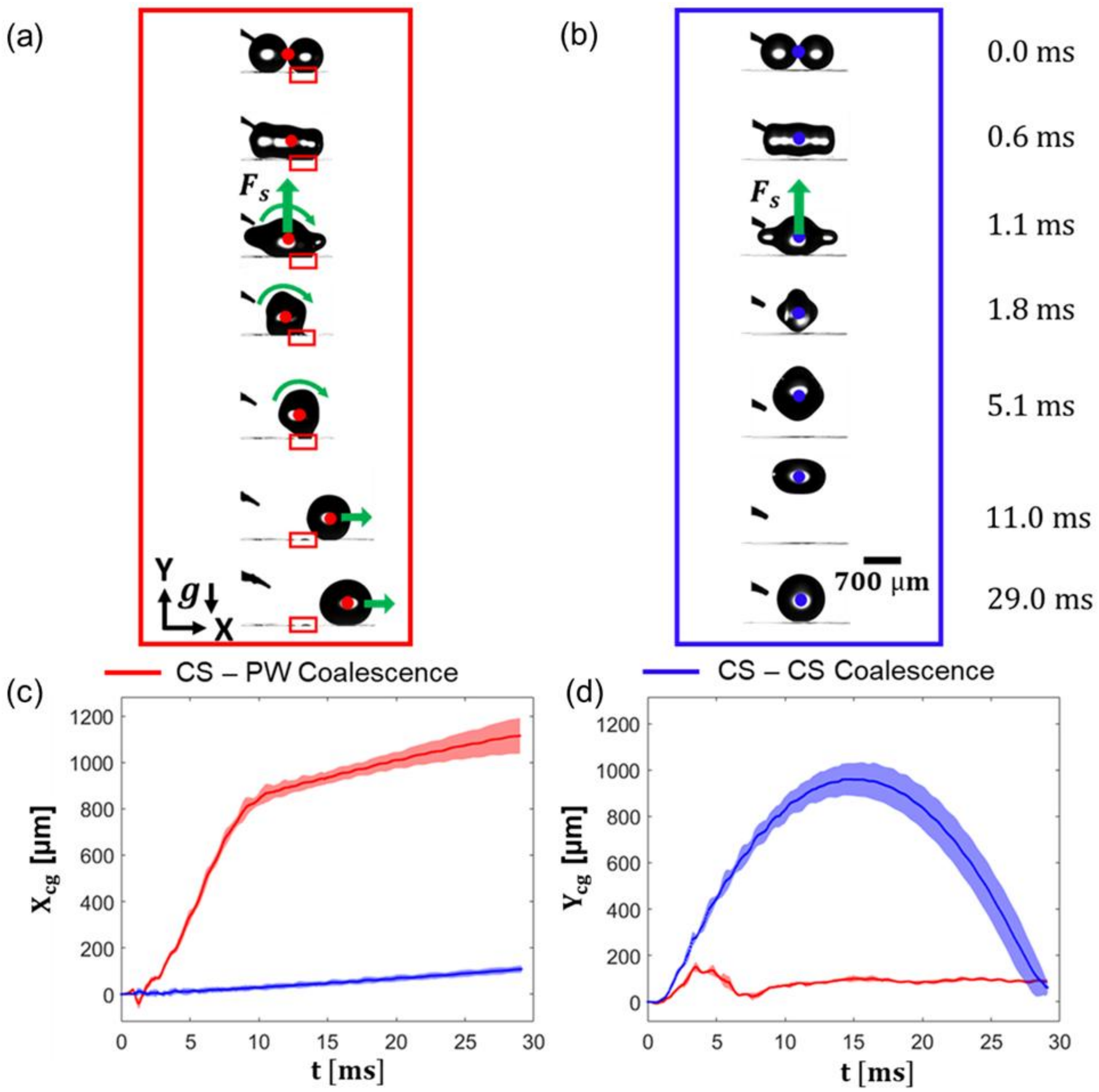

**Figure 2**. (a) High-speed time-lapse images showing coalescence dynamics of Cassie state (CS) and partially wetting (PW) droplets. The left droplet, positioned on the planar region of the superhydrophobic surface, adopts the Cassie state, and the right droplet, positioned on a hydrophilic spot of diameter~ 240 μm, configures as the PW state. The location of the hydrophilic spot, the droplet centroid, surface countering force, the droplet rotation, and the direction of droplet translation are marked with red squares, red circles, green upward arrows, green curved arrows, and green straight arrows, respectively. (b) Coalescence-induced jumping of CS-CS droplets on the superhydrophobic surface, where the droplet centroid is marked with blue circles. (c) And (d) compare the variation of $X_{cg}$ and $Y_{cg}$ with coalescence time (t) for CS-PW droplets (red colour) and CS-CS droplets (blue colour) respectively. The droplet centroid is determined by assuming that the coalescing droplets underwent symmetric deformation about the plane (X-Y plane) of view (the coordinate system is defined in **Fig. 1**).

**Figure 2(a)** depicts that, upon initiation of coalescence between the CS and PW droplets, surface energy is released, generating capillary waves that propagate along the water–air interface. Simultaneously, a liquid bridge forms and advances toward the surface. At approximately 0.6 ms, the left lobe of the merged droplet lifts from the surface, while the right lobe remains pinned at the hydrophilic spot, effectively acting as a fixed pivot point. As the liquid bridge came in contact with the surface, the merged droplet experienced a surface counteracting force $(F_S)$[12] acting at the centroid of the coalescing droplet. This force, applied to the left of the pivot point, induces a net clockwise torque on the asymmetric droplet.

As a result, the coalesced droplet begins to rotate about the pivot point, starting from the left side ($t = 1.1 - 11.0$ ms, **Fig. 2(a)**), and continues rotating until it lands on the right side. The droplet ultimately detaches from the hydrophilic spot at $\sim$ 11.0 ms and transitions into translational motion across the surface, with its position captured up to 29.0 ms. In comparison, coalescence between two CS droplets, where no asymmetric pinning or adhesion force is present, exhibits a symmetric behavior. The merged droplet undergoes vertical, self-propelled jumping: it lifts off from the surface at $\sim$ 1.8 ms, reaches its maximum height at $\sim$ 11 ms, and returns to the surface by $\sim$ 29 ms, as shown in **Fig. 2(b)**. Since the current experiment involved gently guiding the CS droplet into contact with the PW droplet, resulting in purely in-plane motion of the coalesced droplet, an additional experiment is conducted to determine whether the velocity imparted by the needle played any role in producing the observed coalescence behavior. In this control experiment, no needle is used, and the ratio of the PW droplet size to the defect size is kept consistent with the original setup. The results, which are consistent with the main experiment, confirm that the unique coalescence morphology and complete in-plane motion generation are not influenced by needle-induced velocity (refer to Supplementary Material section S1 for further details).

The experimental images are processed using ImageJ software, and the centroid position of the coalescing droplet is tracked over time. The centroid displacement in the x-direction $(X_{cg})$ and in the y-direction $(Y_{cg})$ are plotted as a function of coalescence time $(t)$. As shown in **Fig. 2(c),** for CS -PW droplet coalescence, the slope of the graph remains nearly constant until approximately 10 ms after the onset of coalescence, indicating steady lateral momentum generation before detachment. Subsequently, a marked reduction in slope is observed, consistent with the detachment of the coalesced droplet from the pivot point around 11 ms, as evident in **Fig. 2(a).** The slope of the curve after 11 ms represents the in-plane (lateral) velocity of the droplet post-detachment, which is measured to be approximately 0.03 m/s. In contrast, for CS-CS droplet coalescence, the slope of the curve remains nearly zero, indicating negligible lateral momentum. Instead, the merged CS-CS droplets exhibit out-of-plane jumping behavior. The vertical jumping velocity, extracted from the slope of the corresponding $Y_{cg}$ – time (t) curve at the jumping point (**Fig. 2(b)**), is measured to be approximately 0.15 m/s as shown in **Fig. 2(d).** This marked difference highlights the influence of the hydrophilic anchor, which introduces an additional energy dissipation pathway for the CS-PW droplets coalescence, where a significant portion of the surface energy released during coalescence is dissipated at the pinning spot due to the work of adhesion.[12,19]

To further explore the dynamics of CS-PW droplet coalescence, a series of experiments conducted by varying the sizes of CS ( $d_{cs}$) and PW ( $d_{pw}$) droplets. High-speed video recordings are analyzed using ImageJ to extract the in-plane ($V_{dx}$) and out-of-plane velocity ($V_{dy}$) velocity components, from which the total detachment velocity ($V_d$) and droplet departure angle ($\theta_d$) are calculated to characterize the coalescence-induced jumping behavior (refer to Supplementary Material section S2 for further details).

### A. Equal-size droplet coalescence

First, we explore coalescence between CS and PW droplets of similar size. Here, the droplet sizes are controlled such that the ratio of the sizes of CS and PW droplets $\left(D^* = \frac{d_{pw}}{d_{cs}}\right)$ lies in the range $0.93 \leq D^* \leq 1.07$. The coalescence mechanism of CS-PW droplets can be more effectively described using a non-dimensional size parameter, where the size of the PW droplet is normalized by dividing it by the diameter of the hydrophilic spot ($d_{hs}$). This non-dimensional parameter is defined as $\left(S^* = \frac{d_{pw}}{d_{hs}}\right)$. Using this parameter, the coalescence dynamics are further examined by plotting the velocity and jumping angle of the coalesced droplet as a function of $S^*$. The results revealed a strong dependence of the coalescence dynamics on $S^*$ as shown in **Fig. 3**.

The experimental observations reveal four distinct regimes of coalescence dynamics:

I. **Pinned Coalescence Regime** - The coalesced droplet remains pinned to the hydrophilic spot, with no detachment.

II. **Lateral Detachment Regime** - Detachment occurs with complete in-plane motion.

III. **Oblique Ejection Regime** – The detached droplet ejects at an angle due to a combination of in-plane and out-of-plane motion.

IV. **Normal Ejection Regime** – The detached droplet ejects nearly vertically from the surface (complete out-of-plane motion).

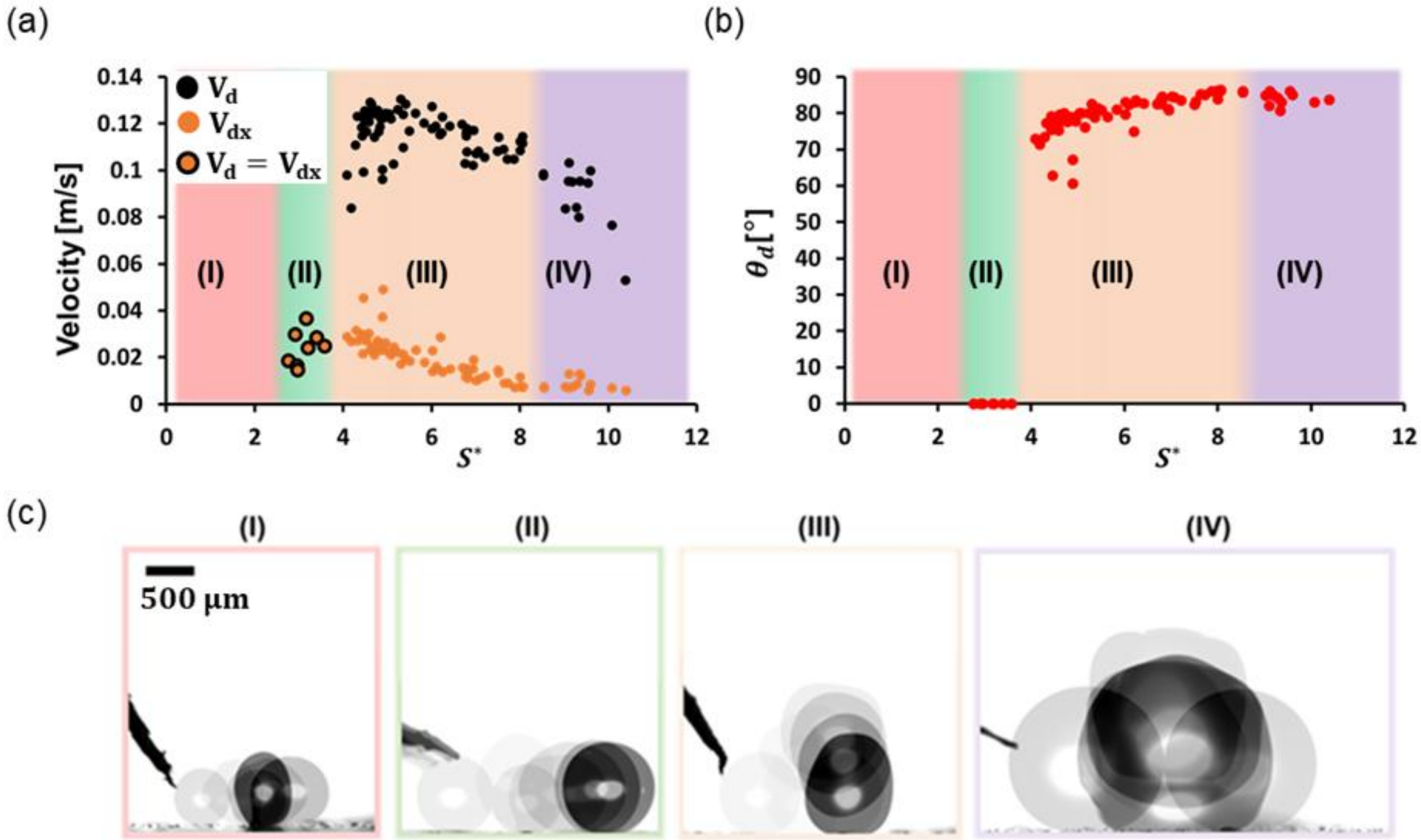

**Figure 3.** Dynamics of coalescence between CS and PW droplets of the same size: (a) Coalescence-induced droplet total velocity ($V_d$) and in-plane velocity ($V_{dx}$) after detachment from the hydrophilic spot as a function of $S^*$. (b) Droplet departure direction $\theta_d$ as a function of $S^*$.The transition zone between the different regimes is not sharp. (c) Chronophotography images of typical coalescence events for (I) the pinned coalescence regime, (II) the lateral detachment regime, (III) the oblique ejection regime, and (IV) the normal ejection regime are shown at intervals of 2 ms, 6 ms, 5.6 ms, and 5 ms, respectively, to illustrate droplet motion in different regimes. In each case, the first image is the brightest and the last image is the darkest.

The pinned coalescence regime (I) is found for $S^*$ values below ~3 and a narrow region of the lateral detachment regime (II), with $V_d = V_{dx}$ and $\theta_d$~0°, is found for $S^*$ between ~3 to ~3.5. The oblique ejection regime (III), with $\theta_d < 85°$, for $S^*$ between ~4 to ~8, and afterwards the normal ejection regime (IV) exists, where $V_{dx} \sim 0$ and $\theta_d$~90°.

**Figure 4** illustrates the temporal evolution of coalescence dynamics across the four identified regimes. In the pinned coalescence regime (I) shown in **Fig. 4(a),** the coalesced droplet undergoing asymmetrical coalescence, experiences a clockwise torque, and due to that, the drop rotates and lands on the right side, bounces, and again reverses its rotational direction. The droplet continues to swing back and forth about the pinning location until the rotational kinetic energy is fully dissipated through viscous losses. In the lateral detachment regime (II) shown in **Fig. 4(b),** the rotating coalesced asymmetrical shaped droplet detaches from the pinning spot and translates across the surface as described in the explanation for **Fig. 2(a)**.

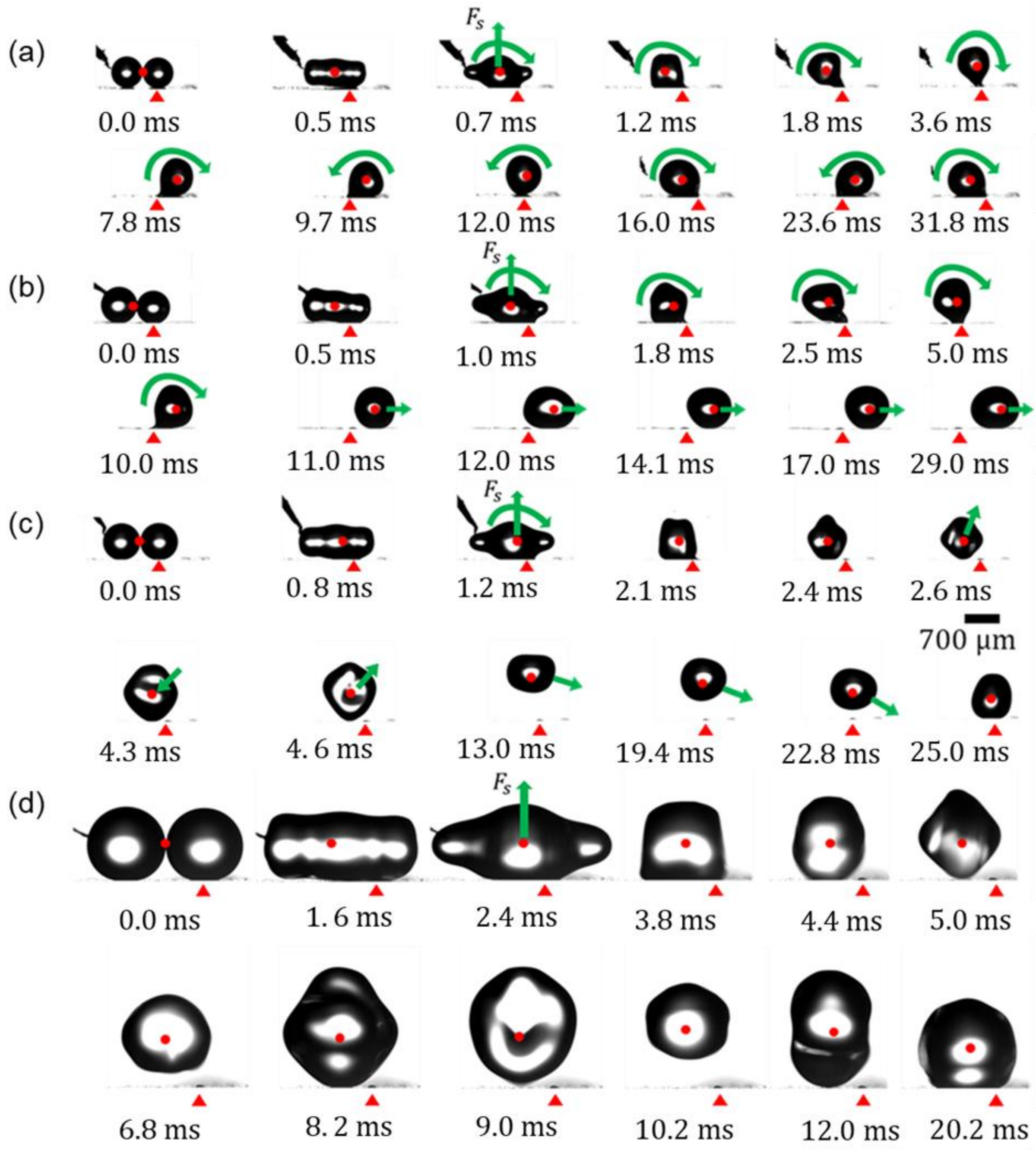

**Figure 4.** Compares the high-speed time-lapse images of coalescence dynamics for all four regimes of equal-sized CS-PW droplets coalescence. (a) Pinned coalescence regime, (b) lateral detachment regime, (c) oblique ejection regime, and (d) normal ejection regime. The location of the hydrophilic spot, the droplet centroid, surface countering force, the droplet rotation, and the direction of droplet translation are marked with red upward arrowheads, red circles, green upward arrows, green curved arrows, and green straight arrows, respectively. The droplet centroid is determined by assuming that the coalescing droplets underwent symmetric deformation about the plane of view.

In the oblique ejection regime (III) shown in **Fig. 4(c),** the coalesced droplet does not exhibit significant rotation. Instead, it detaches from the hydrophilic spot at approximately 2.4 ms, performs its first jump at around 2.6ms, briefly recontacts the surface, and then jumps again at approximately 4.6 ms. During this process, the asymmetric curvature of the droplet interacts with the surface, resulting in an inclined reaction force that propels the droplet in an oblique direction. After ejection, the droplet typically lands again on another hydrophilic spot, where it becomes pinned and undergoes damped oscillations. The residual kinetic energy is gradually dissipated through viscous damping. Finally, in the normal ejection regime (IV) shown in **Fig. 4(d)**, the surface energy released during coalescence is high enough to render the effect of pinning negligible. The coalescence becomes nearly symmetrical, and the detached droplet ejects vertically in a fully out-of-plane motion. This behavior closely resembles that observed during symmetrical CS–CS droplet coalescence.

While the experimental investigation provides valuable insights into the morphological evolution across different coalescence regimes, it lacks the resolution to capture detailed hydrodynamic phenomena, specifically, the spatial and temporal distributions of velocity and pressure fields arising from interfacial curvature changes during coalescence. These intricate dynamics remain largely inaccessible to experimental observation, particularly at the microscale, where rapid interface motion predominates. To overcome this limitation, we employ a three-dimensional numerical modeling framework based on the Volume of Fluid–Continuum Surface Force (VOF-CSF)[20–22] method, incorporating with static contact angle (SCA) model to simulate the coalescence of same-sized CS-PW droplets (refer to Supplementary Material section S3 for further details). Similar to the experiments, the numerical model also successfully captures the four distinct coalescence regimes (I–IV). However, some differences in the transition boundaries are observed. In the experiments, the pinned coalescence regime (Regime I) occurs for $S^*$ values below 3, whereas in the numerical investigation, this regime was observed for $S^*$ value below 4.1. The lateral detachment regime (Regime II) is found for $S^*$ between ~ 3 to ~ 3.5 in the experiments, while in the numerical simulation, the regime II is found for $S^*$ between ~ 4.3 to ~ 4.2. Unlike the experiments, where the oblique ejection regime (Regime III) was identified for $S^*$ between ~ 4 to ~8, the simulations revealed this regime, with a jumping angle, with $\theta_d < 85°$, for a wider range of $S^*$ between ~ 5 to ~12. Beyond this range, the normal ejection regime (Regime IV) is observed, where $V_{dx} \sim 0$ and $\theta_d \sim 90°$.

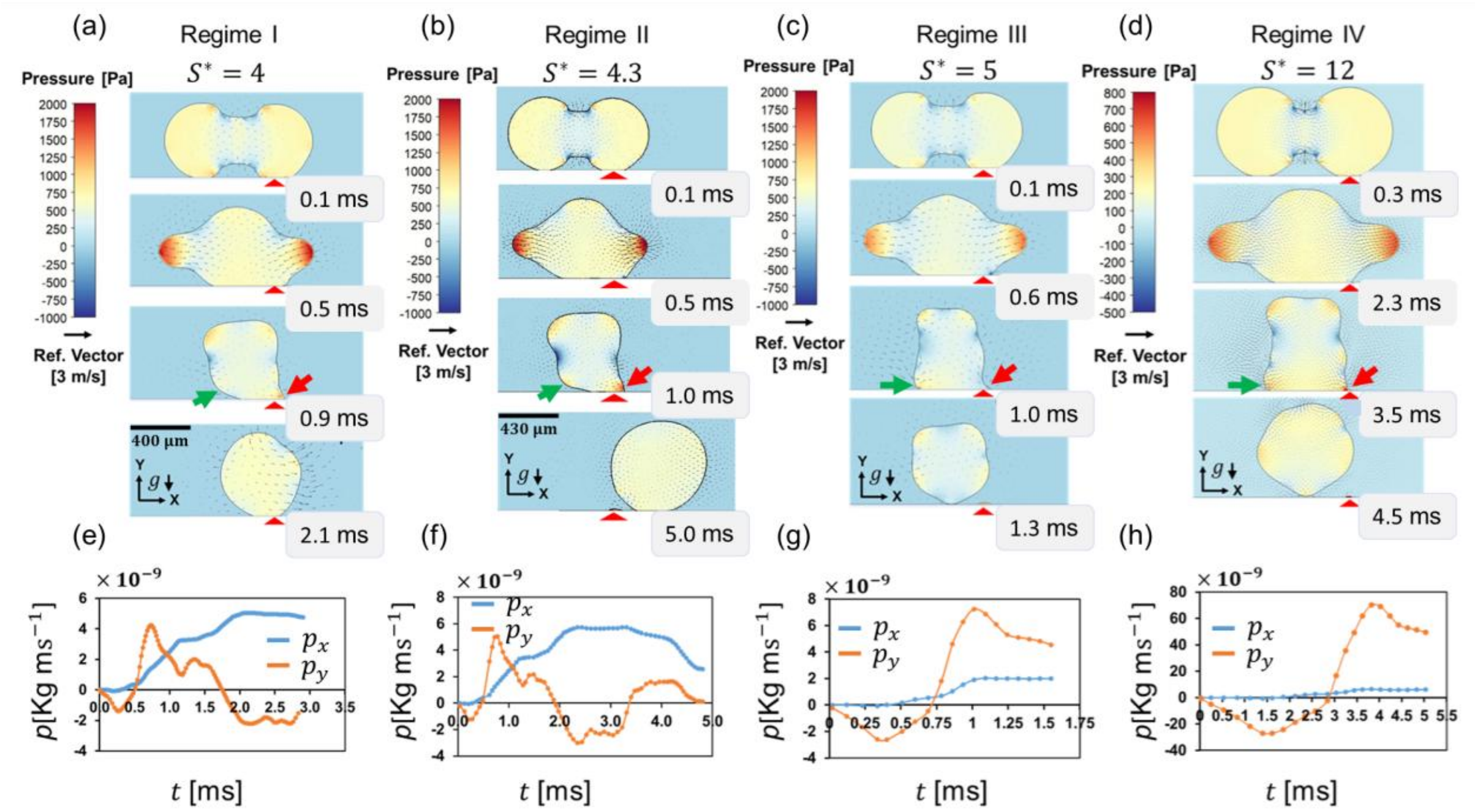

**Figure 5.** Temporal evolution of velocity and pressure fields at the symmetry plane during coalescence of same-sized CS–PW droplets: (a) 400 µm droplets in the pinned coalescence regime (Regime I), (b) 430 µm droplets in the lateral detachment regime, (c) 500 µm droplets in the oblique ejection regime (Regime III), and (d) 1200 µm droplets in the normal ejection regime (Regime IV). The superhydrophilic spot is marked by a red upward arrow. Corresponding temporal variations of in-plane momentum ( $p_x$) and out-of-plane momentum ( $p_y$) are shown in (e), (f), (g), and (h), respectively.

The hydrodynamics, along with the associated variation in in-plane momentum ( $p_x$) and out-of-plane momentum ( $p_y$) are compared for each of the coalescence dynamics regimes as illustrated in **Fig. 5**. In Regime I, as shown in **Fig. 5(a)**, immediately after the onset of coalescence, a low-pressure liquid bridge forms between the CS and PW droplets, facilitating fluid motion and bridge expansion, as observed at 0.1 ms. The merging droplet experiences a pinning force at the right lobe, resulting in an asymmetric liquid body. This asymmetry leads to an uneven distribution of pressure and velocity fields within the flow domain. By time 0.5 ms, the velocity vectors become more inclined, with the flow moving downward in the right lobe compared to the left. At 0.9 ms, the left lobe lifts off from the wall, where a positive pressure region forms (indicated by the green arrow), while a negative pressure region begins to develop near the pinned edge (indicated by the red arrow). In this regime, a significant amount of $p_x$ is generated and becomes maximum ~2.1 ms as shown in **Fig. 5(e).** The pinned droplet configuration corresponding to this peak $p_x$ is shown at 2.1 ms.

The hydrodynamics, along with the associated variation in $p_x$ and $p_y$ for the lateral detachment regime (II), are illustrated in **Figs. 5(b,f)**. The coalescence process progresses similarly to Regime I. However, after 5 ms, distinct differences emerge. Unlike the pinned case, the coalesced droplet detaches from the pinning site and moves laterally along the surface without jumping.

The hydrodynamics, along with the associated variation in $p_x$ and $p_y$ for the oblique ejection regime (III), are illustrated in **Figs. 5(c,g)**. The coalescence process initially progresses similarly to Regime II. However, after 0.6 ms, distinct differences emerge. In comparison to the pinned case, the curvature of the merged droplet evolves such that a negative pressure region develops near the pinned edge (indicated with a green arrow), while a positive pressure region forms at the opposite edge (indicated with a green arrow), as observed at 1.0 ms. The $p_y$ becomes maximum values at this instant, as shown in **Fig. 5(g)**. Afterwards, the merged droplet detaches from the pinning spot. At the moment of the jump, at 1.3 ms, the pressure distribution within the droplet becomes highly asymmetrical, and the droplet departs obliquely from the wall. The hydrodynamics and associated variation in $p_x$ and $p_y$ for the normal ejection regime (IV) are illustrated in **Figs. 5(d,h)**. In this case, the pressure and velocity variation shows a nearly symmetrical variation in the left and right lobes of the liquid body. The droplet morphology is found to be similar to that of Regime III, just before detachment of the liquid body from the pinning spot, as shown at 3.5 ms.

The momentum variation plots reveal that in Regimes I and II, both in-plane momentum ($p_x$) and out-of-plane momentum ($p_y$) are generated, with $p_x$ being larger than $p_y$, as shown in **Fig. 5(e,f)**. In the oblique ejection regime (Regime III), the generation of $p_x$ is smaller than $p_y$ as shown in **Fig. 5(g)**. For the case of the normal ejection regime (Regime IV), the generation of $p_x$ becomes negligible compared to $p_y$ as shown in **Fig. 5(h)**.

To investigate the hydrodynamic variations in the absence of a pinning spot during the coalescence of equal-sized droplets, simulations were conducted for CS–CS droplet coalescence, using the same droplet size as in the CS–PW coalescence case (as shown in **Fig. 5**). The results demonstrate symmetrical coalescence behavior across all cases, with negligible generation of in-plane momentum (refer to **Fig.S.8**, Supplementary Material subsection S3.3).

## B. Effect of size mismatch between CS and PW droplets on the coalescence-induced lateral momentum generation

To examine the impact of size mismatch on jumping velocity and direction, additional experiments are conducted using droplets of unequal radii, a scenario that more accurately represents typical condensation conditions.[23] Here $S^*$ is fixed and the size of the CS droplet, $(d_{cs})$ is varied. The resulting jumping velocities and angles are measured and are presented in **Fig. 6,** which illustrates the transition between coalescence regimes based on size mismatch. The oblique ejection regime (III) is observed for $0.5 < D^* < 0.8$, the lateral detachment regime appears for $0.9 < D^* \leq 1$, and the pinned coalescence regime is found for $1 < D^*$.

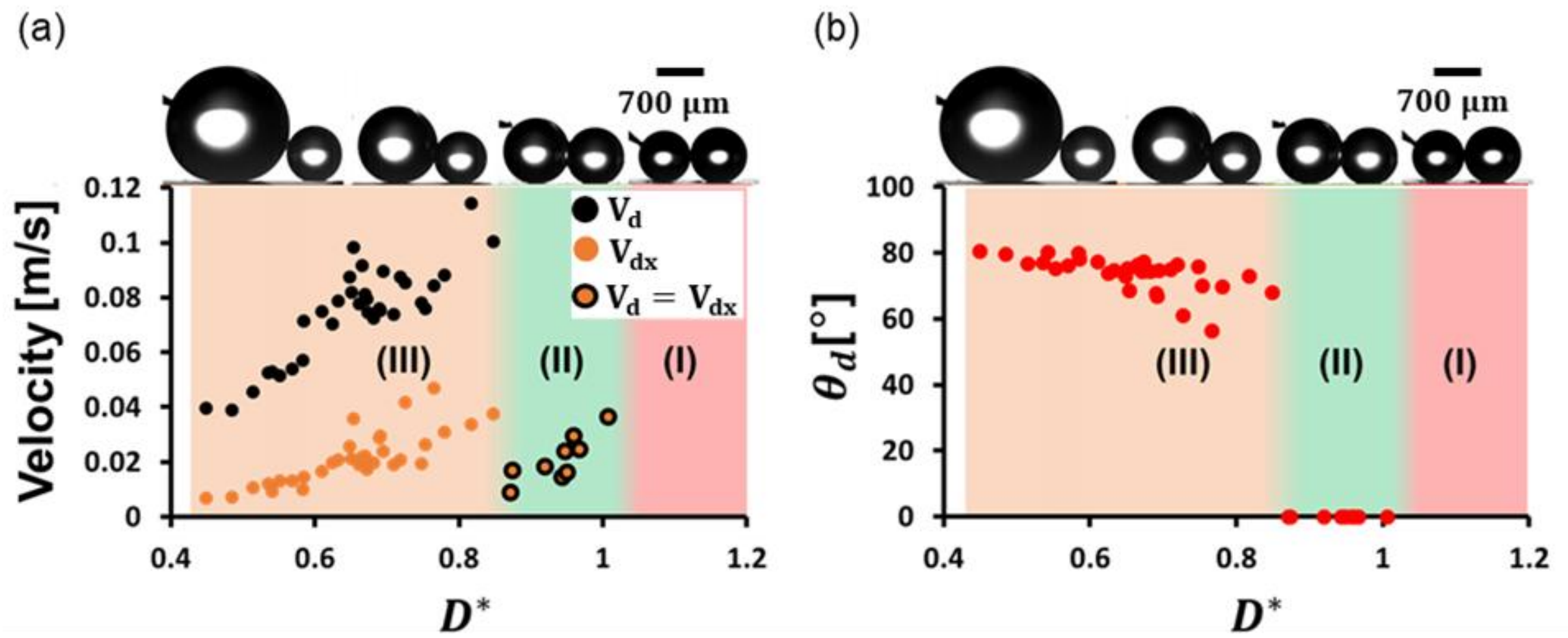

**Figure 6.** Unequal-size Cassie state (CS) and partially wetting (PW) droplets coalescence. Where the size of the PW droplet $S^* \sim 3.25 \pm 0.25$, and the size of the CS droplet is varied. (a) Droplet total velocity $V_d$ and the lateral velocity $V_{dx}$ generated in the experiment at the point of detachment as a function of $D^*$. (b) Droplet jumping direction at the time of detachment from the surface as a function of $D^*$. The transition zone between the different regimes is not sharp.

The detailed coalescence dynamics of mismatch sizes CS-PW droplets are described in **Fig. 7.** Notably, for extreme size mismatch, where $D^* < 0.5$, a unique regime is observed as shown in **Fig. 7(a).** Upon coalescence, the merged droplet forms a highly asymmetric shape, with the mass distributed between a larger left lobe and a smaller right lobe that remains pinned. Unlike other regimes, the left lobe does not lift off the surface. Instead, the entire merged droplet translates laterally across the surface, eventually reaching the pinning site. There, it becomes pinned and undergoes oscillatory motion until its energy is dissipated through viscous losses.

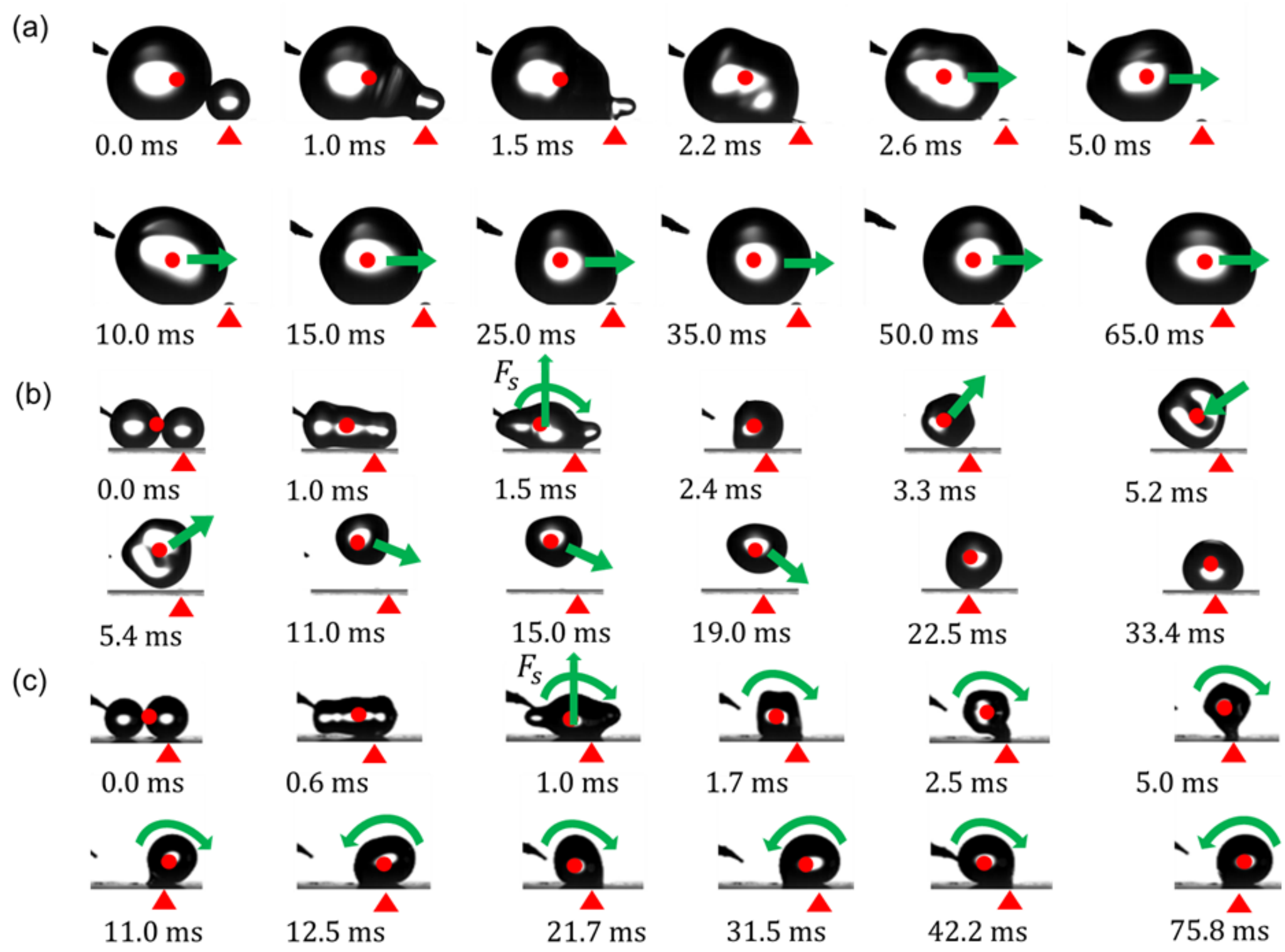

**Figure 7.** Compares the high-speed time-lapse images of coalescence dynamics for unequal sizes of CS-PW droplets. (a) Coalescence dynamics during $D^* < 0.5$, the left lobe of the asymmetric droplet does not lift the surface at $\sim$ 1ms compared to other regimes of coalescence dynamics. (b) Oblique ejection regime and (c) pinned coalescence regime. The location of the hydrophilic spot, the droplet centroid, surface countering force, the droplet rotation, and the direction of droplet translation are marked with red upward arrow heads, red circles, green upward arrows, green curved arrows, and green straight arrows, respectively. The droplet centroid is determined by assuming that the coalescing droplets underwent symmetric deformation about the plane of view.

To understand the hydrodynamics of unequal-sized CS–PW droplet coalescence, numerical simulations were performed. Similar to the experimental observations, the numerical simulations capture both the pinned coalescence regime (Regime I) and the oblique ejection regime (Regime III) and coalescence dynamics at extreme size mismatch $D^* = 0.5$. The pinned regime is observed at $D^* = 1.25$, while the oblique ejection regime occurs at $D^* = 0.83$, which is consistent with the experimental findings described in **Fig. 6**. The pressure and velocity variation shows for unequal-sized CS-PW droplets coalescence, similar to the equal-sized CS-CS droplets coalescence (refer to Supplementary Material subsection S3.4).

## CONCLUSION

This study provides a comprehensive experimental investigation into the coalescence dynamics of Cassie state (CS) and partially wetting (PW) droplets on a nanostructured superhydrophobic surface. A droplet placed directly on the superhydrophobic surface forms a CS, while a droplet deposited on a hydrophilic spot patterned onto the globally superhydrophobic surface assumes a PW state. Using high-speed imaging, we identified four distinct coalescence regimes—pinned coalescence, lateral detachment, oblique ejection, and normal ejection—governed primarily by the non-dimensional size of the PW droplet ($S^*$) for equal-sized droplets coalescence. For the unequal-size droplets coalescence scenarios, the normal ejection regime is not observed. Notably, at extreme droplet size mismatch ($D^* < 0.5$), a unique in-plane translational motion of the merged droplet is detected. An experimentally validated VOF-based numerical modeling framework is utilised to investigate the hydrodynamics of coalescence between Cassie state (CS) and partially wetting (PW) droplets. Through this study, three distinct coalescence regimes are captured for equal-sized CS–PW droplet configurations, namely, pinned coalescence, oblique ejection, and normal ejection. Consistent with the experimental outcomes, the normal ejection regime is not observed in the numerical simulations for unequal-size CS-PW droplets coalescence. Notably, at extreme size mismatch $D^* = 0.5$, a unique in-plane translational motion of the merged droplet is observed, which is also observed in the experiment. The criterion for droplet detachment from the pinning spot is identified by a change in interfacial curvature: just before detachment of the merged droplet, the curvature at the pinning spot becomes negative, while the curvature on the opposite side becomes positive. These findings offer new insights into the asymmetric coalescence dynamics of sessile droplets interacting across two extreme wetting states and highlight the critical influence of droplet size mismatch on the resulting motion and regime transitions.

## ACKNOWLEDGEMENTS

We gratefully acknowledge the financial support provided by IIT Ropar through ISIRD (Grant no. 9-388/2018/IITRPR/3335) for this work. We also acknowledge valuable discussions with Anand S and Varun Chaturvedi of Thermofluidics Research Laboratory, IIT Ropar on various aspects of droplet dynamics investigated in this study.

## AUTHOR DECLARATIONS

### Author Contributions

C.S.S. conceived and supervised the research and arranged the funding. C.S.S. and R.K. designed the experiments, R.K. fabricated the samples. R.K. and G.C.P performed the experiments. R.K. conducted the numerical simulations. R.K., G.C.P., and C.S.S. analyzed the results. All authors contributed to writing the manuscript.

**Competing interest**

All authors declare they have no competing interests.

**Additional Information**

Supplementary materials are provided separately with this manuscript.

**Data Availability**

The data that support the findings of this study are available within the article and its supplementary material.

# Supplementary Material

# Pinning-mediated coalescence-induced lateral droplet motion on a nanotextured superhydrophobic surface

**Raushan Kumar, Gopal Chandra Pal, Chander Shekhar Sharma***

Thermofluidics Research Laboratory, Department of Mechanical Engineering, IIT Ropar, Punjab-140001, India

*Email: chander.sharma@iitrpr.ac.in, Ph: +91-1881-232358

**TABLE OF CONTENTS**

**Section S1: Self-induced Cassie state (CS) and partially wetting (PW) droplets coalescence**

The experimental result reveals a purely in-plane motion of the coalesced droplet during CS–PW coalescence, as shown in **Fig. 4.2(a)** of the main text, where a superhydrophobic wooden needle is used to initiate coalescence. To examine whether the needle imparts additional energy influencing this unique in-plane motion, the present investigation is conducted without any external triggering mechanism, aiming to identify the conditions under which lateral detachment occurs purely due to spontaneous coalescence of a Cassie-state (CS) droplet and a partially wetting (PW) droplet, free from external disturbance.

To perform this new type of experiment, a specially designed surface was fabricated, followed by a series of controlled coalescence experiments.

**Subsection S1.1: Surface fabrication**

One strip of Al-6061 (1 mm thickness) (Goodfellow Cambridge Ltd, UK) is used as the base surface, a hole is drilled in the middle of the sample of size $\sim$ 100 μm drill bit using the Hybrid micromachining centre- DT110i (Mikrotools Pte Ltd.). Afterwards, the strips underwent a thorough cleaning process involving Isopropanol (Sigma-Aldrich) and deionised (DI) water rinse. One face of the cleaned strips is super-hydrophobised by spray coating of a commercially available solution of silica nanoparticles, homogenously dispersed in isopropanol solution (Glacco Mirror coat "Zero", Soft99.)[16]. Then, the contact angle variation is measured on the coated surface using a goniometer (Holmarc Opto-Mechatronics Ltd, Model No- HO-IAD-CAM-01A-01A, India). The apparent advancing contact angle ($\theta_{ACA}$) and contact angle hysteresis (CAH) are obtained as $163° \pm 1.8°$ and $7° \pm 1.2°$, respectively. The central hole in the surface is used to generate and retain a droplet in the partially wetting (PW) state on the globally superhydrophobic surface.

**Subsection S1.2: Experimental Outcome**

**Figure S.1** shows the schematic diagram of the experimental set-up used for self-induced CS-PW droplets coalescence. In this setup, a CS droplet is dispensed near the central hole using in-house developed superhydrophobic micropipette tips. Simultaneously, a PW droplet is generated through the central hole using a syringe pump (New Era), connected to a 22-gauge dispensing needle (Hamilton) with an outer diameter of 718 μm at a controlled flow rate of 1 $\mu l/\mathrm{min}$. As the PW droplet grows and eventually contacts the CS droplet, asymmetric

coalescence is initiated. This interaction closely resembles the behavior described in **Fig. 4.2(a)** of the main text.

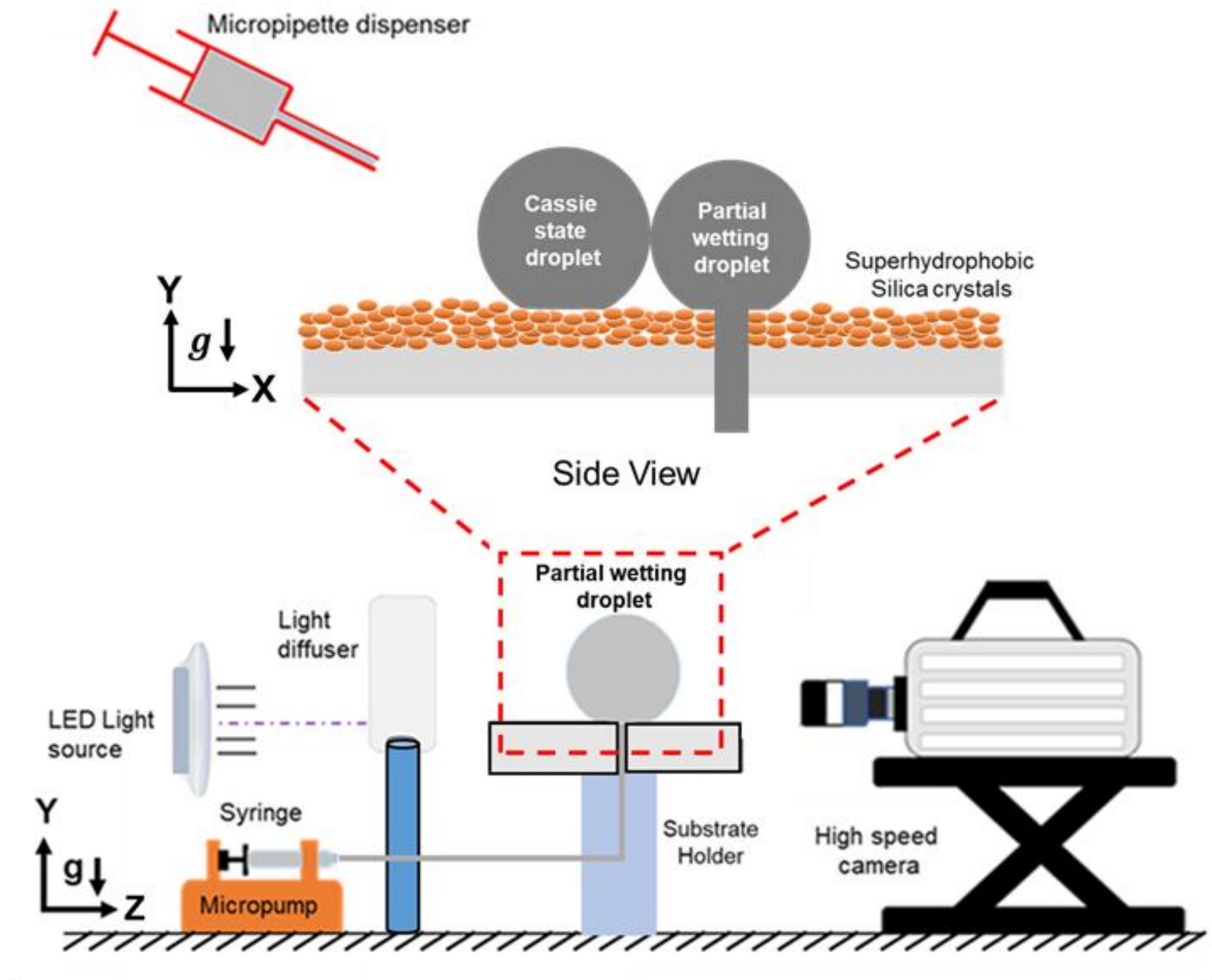

**Figure S.1.** Experimental setup for self-induced coalescence of Cassie state (CS) and partially wetting (PW) state droplets. Where droplets are coalescing with any external triggering mechanism.

The pre-coalescence state (~66 ms before onset, **Fig. S.2**) is captured, showing the CS and PW droplets just before merging. After the onset of coalescence, the merging droplets exhibit asymmetric behavior, with a liquid bridge forming at the point of contact and growing over time ($t = 0.0 - 0.7$ ms, **Fig. S.2**). Once this bridge makes contact with the surface, the newly formed asymmetric droplet, pivoted on the liquid column beneath the hole, experiences a net surface countering force ($F_S$) acting at its center of gravity (cg). Since this force acts to the left of the pivot point, it generates a net clockwise torque on the droplet. As a result, the coalesced droplet begins to rotate about the pivot point, starting from the left side ($t = 0.7 - 6.5$ ms, **Fig. S.2**), and continues rotating until it lands on the right side. The droplet ultimately detaches from the liquid column at approximately 6.5 ms and transitions into translational motion across the surface, with its position recorded at 40.5 ms. This experiment confirms that self-induced coalescence between CS and PW droplets can drive complete in-plane motion on a globally superhydrophobic surface, with capillary forces overcoming the pinning resistance and generating a surface countering force-induced torque.

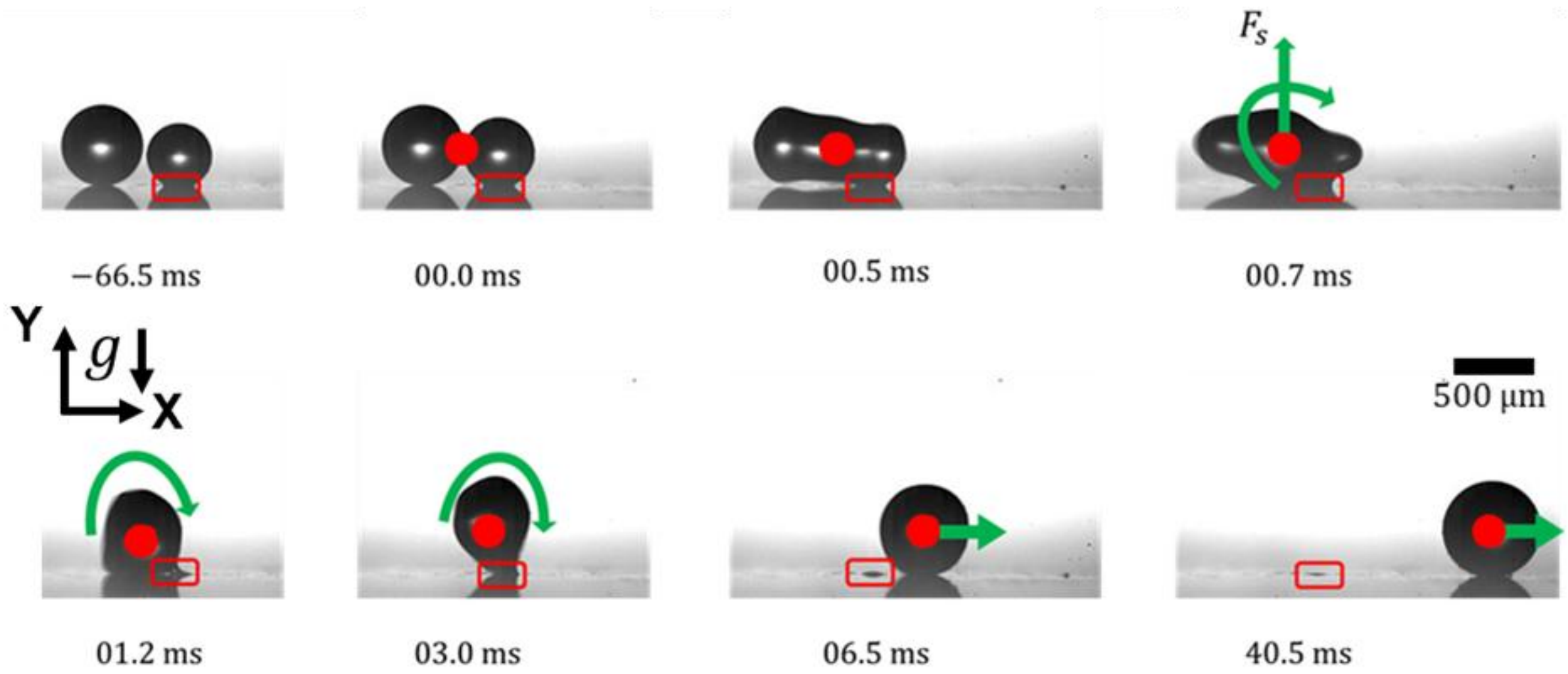

**Figure S.2.** Self-induced coalescence dynamics of a Cassie state (CS) and partially wetting (PW) droplet. The left droplet, positioned on the planar region of a superhydrophobic surface, adopts the Cassie state. The right droplet, generated through a central hole in the surface, is connected with a liquid column of diameter ~ 160 µm, which configures a partially wetting state. The location where the droplet connected to the liquid column, the droplet centroid, surface countering force, the droplet rotation, and the direction of droplet translation are marked with red squares, red circles, green upward arrows, green curved arrows, and green straight arrows, respectively. The droplet centroid is determined by assuming that the coalescing droplets underwent symmetric deformation within the plane (X-Y plane) of view (the coordinate system is defined in **Fig. S.1**).

**Section S2: Droplet detachment velocity and direction calculation**

High-speed videos are analyzed using ImageJ to extract kinematic parameters. The droplet centroid is tracked frame by frame, and the center of gravity (cg) is plotted as a function of time t at the moment of detachment from the hydrophilic spot. The variation of $X_{cg}$ (in-plane direction) showed a linear trend, where the slope corresponds to the in-plane velocity ($V_{dx}$) of the coalesced droplet as shown in **Fig. 4.3 (a)**. The variation of $Y_{cg}$ (out-of-plane direction) followed a parabolic fit, and the slope of the curve at the detachment point is used to determine the out-of-plane velocity ($V_{dy}$) as shown in **Fig. 4.3 (b)**. The total velocity ($V_d$) of the droplet post-detachment is then computed as the resultant of these two components. The jumping direction, calculated by the droplet departure angle ($\theta_d$), is obtained through $tan(\theta_d) = \frac{V_{dy}}{V_{dx}}$.

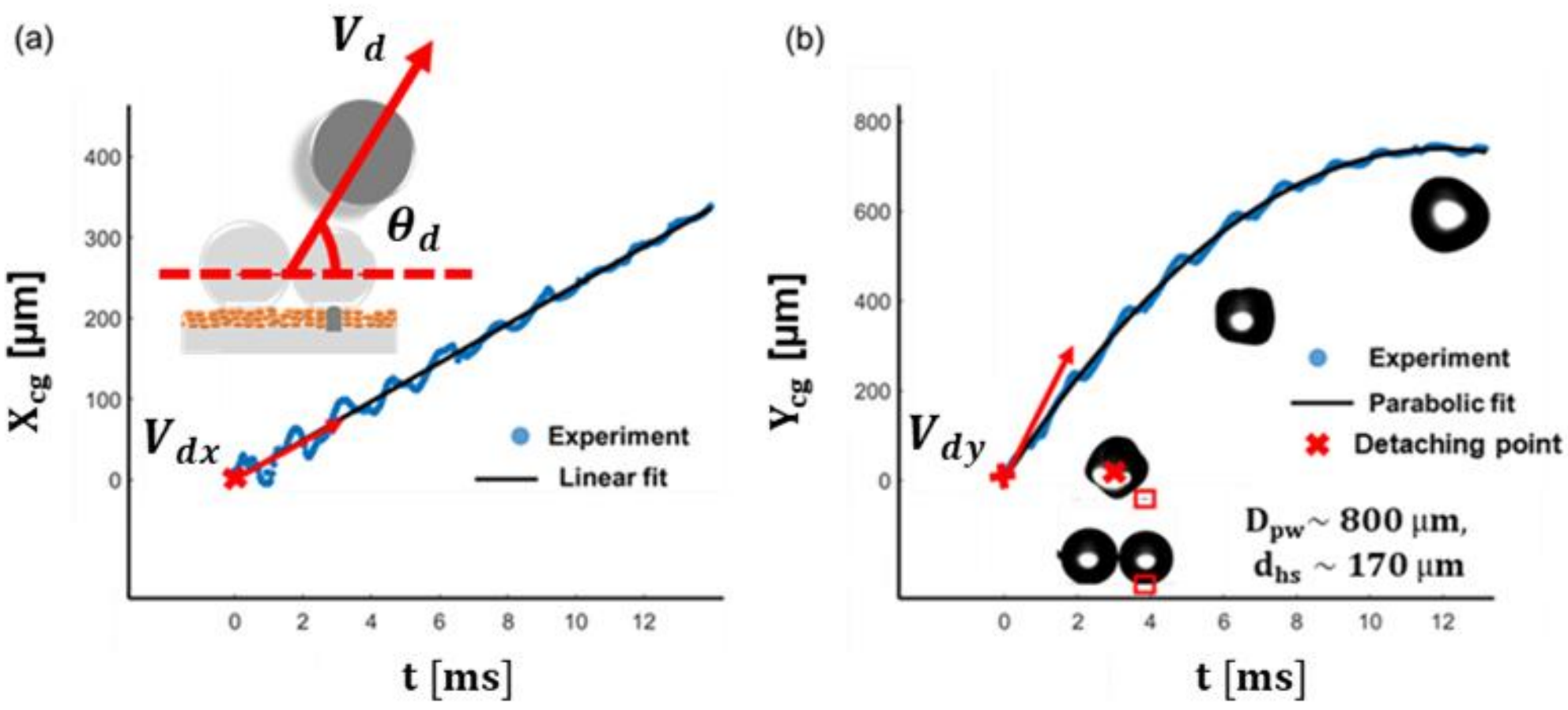

**Figure S.3.** Displacement of the droplet centroid in the (a) x-direction and (b) y-direction as a function of time t. The CS and PW droplets have an initial size of ~ 800 μm, and the hydrophilic spot size ( $d_{hs}$ ) is ~ 170 μm. The inset image shows the definition of departure angle ($\theta_d$).

## Section S3: Numerical investigation of CS-PW droplets coalescence

### Subsection S3.1: Computational model and validation

The Volume of Fluid–Continuum Surface Force (VOF-CSF)[24–28] method, incorporating with static contact angle (SCA) model, is used here to simulate the coalescence of CS-PW droplets. This method models non-diffusive multiphase fluid flow problems by tracking the volume fraction of each phase of the fluids in the flow domain and solving a set of momentum equations. The continuity equation (S1) and Navier–Stokes equation (S2), as written below, are used to govern the flow variables.

$$\frac{\partial \rho}{\partial t} + \boldsymbol{\nabla}.(\rho \boldsymbol{u}) = 0 \tag{S1}$$

$$\frac{\partial (\rho \boldsymbol{u})}{\partial t} + \boldsymbol{\nabla}.(\rho \boldsymbol{u} \otimes \boldsymbol{u}) = -\boldsymbol{\nabla} p + \mu \boldsymbol{\nabla}^2 u + \boldsymbol{F}_{st} + \rho \boldsymbol{g} \tag{S2}$$

Where, $u$, $\rho$, $\mu$, $g$, and $F_{st}$ represent the velocity, density, and dynamic viscosity of the mixture, respectively, and $g$, $F_{st}$ represents the gravitational acceleration and volumetric force at the interface of liquid-vapor resulting from surface tension. In the case of two-phase problems, there is a chance that a particular control volume may occupy multiple phases simultaneously. If $\mu_g$ , $\mu_l$ represents the dynamic viscosity of air and liquid, and $\rho_g$ and $\rho_l$ represent the density of air and liquid, then the apparent viscosity $\mu$ and density $\rho$ in each cell are calculated as:

$$\mu(x,t) = \mu_g + (\mu_l - \mu_g)\alpha \tag{S3}$$

$$\rho(x,t) = \rho_g + (\rho_l - \rho_g)\alpha \tag{S4}$$

Where $\alpha$ is the ratio of cell volume occupied by the liquid to the total volume of the control cell. The numerical value of $\alpha$ in a cell lies between 1 and 0. If a cell is filled with liquid, then $\alpha = 1$, for any cell filled with air, $\alpha = 0$, and if a cell contains an interface between air and water, then $0 < \alpha < 1$. The liquid-vapor interface is tracked by solving the continuity equation for the liquid volume fraction:

$$\frac{\partial \alpha}{\partial t} + \boldsymbol{u}.(\boldsymbol{\nabla}\alpha) = 0 \tag{S5}$$

The surface tension effects at the liquid-vapor interface are modeled using the Continuum Surface Force (CSF) model.[29] It accounts for the surface tension force as a volume force in equation (S2) as:

$$\boldsymbol{F}_{st} = \frac{\gamma_{lv}\rho\kappa\boldsymbol{\nabla}\alpha}{\frac{1}{2}(\rho_l+\rho_g)} \tag{S6}$$

Where the surface tension between the air and water interface is $\gamma_{lv}$ and κ is the curvature of the interface.

$$\kappa = -\boldsymbol{\nabla}.(\boldsymbol{n}) \tag{S7}$$

Where $\boldsymbol{n}$ is the unit normal vector and calculated as $\boldsymbol{n} = \left(\frac{\boldsymbol{\nabla}\alpha}{|\boldsymbol{\nabla}\alpha|}\right)$. At the contact line, $\boldsymbol{n}$ is calculated as:

$$\boldsymbol{n} = \boldsymbol{n_w}\cos(\theta) + \boldsymbol{n_t}\sin(\theta) \tag{S8}$$

Where, $\boldsymbol{n_w}$ and $\boldsymbol{n_t}$ are the unit normal and tangential vectors at the wall directed into the liquid and wall interface.

Experimental observations indicate that, after the coalescence of Cassie state (CS) and partially wetted (PW) droplets, a residual liquid film remains on the hydrophilic spot, as shown in **Fig. S.4**.

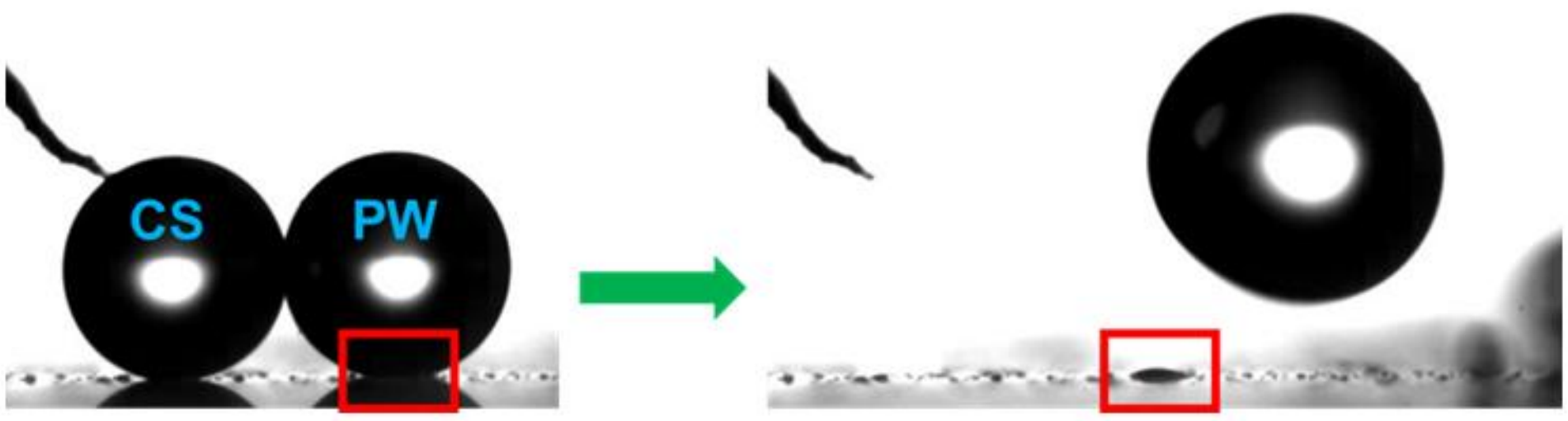

**Figure S.4.** CS–PW droplets at the onset of coalescence and the resulting merged droplet after coalescence. A residual liquid film remains at the hydrophilic spot, highlighted by a red rectangle.

The formation of this pinned liquid film indicates high wettability of the hydrophilic spot. As a result, it can be assumed that the PW droplet effectively behaves as if it is resting on a local superhydrophilic defect. To replicate this condition in the numerical model, a PW droplet is initialized in the computational domain such that its base experiences superhydrophilic wettability. This is achieved by assigning suitable wall boundary conditions within the simulation setup.

**Figure S.5(a)** illustrates the three-dimensional computational domain and the boundary conditions used for simulating the coalescence of CS and PW microdroplets. Experimental observations show that the coalescence dynamics of the CS-PW droplets are symmetric about the mid-plane of the liquid body. Therefore, only half of the coalescing domain is modeled to reduce computational cost. A symmetry boundary condition is applied along the *X–Y* plane to reflect this assumption. The velocity inlet boundary is specified with zero velocity, based on the assumption that the gas-phase velocity is negligible at locations sufficiently far from the liquid–gas interface. A gauge pressure of zero is imposed at the pressure outlet boundary. To simulate the experimentally observed liquid film formation, a circular superhydrophilic region is defined at the wall within the computational domain where the PW droplet is placed. The static contact angle ($\theta_{SCA}$) is used to define the wall wettability. The wall is globally modeled as superhydrophobic with $\theta_{SCA} = 180°$, except for a circular superhydrophilic spot of diameter $d_{sh} = 100$ µm, where $\theta_{SCA} = 0°$ .

The model is validated against experimental results for the equal-sized CS–PW droplets coalescence. Numerical simulations are performed for the same scenario, where the non-dimensional size of the PW droplet is $S^* = 7$. This non-dimensional size is defined as: $S^* = \frac{d_{pw}}{d_{sh}}$ , where $d_{pw}$ represents the diameter of the PW droplet. To check the grid independence, three different meshes are used to perform numerical simulation as shown in **Fig. S.5(b)**. The coarse mesh consists of a larger mesh size of 20 µm in the air domain, while within the liquid domain, the mesh element is reduced by half using a single level of refinement. The fine mesh consists of a larger mesh size of 20 µm in the air domain; however, in the liquid domain, the mesh element is reduced by one-fourth using a double level of refinement. In the finest mesh, the air domain again uses a 20 µm mesh size, with the liquid domain refined by one-fourth, as in the fine mesh. Additionally, at the liquid–air interface, the mesh is further refined to half the size of the mesh in the liquid domain. Following mesh generation, an adaptive meshing technique is applied at the liquid–air interface. This technique relies on the volume fraction gradient of the liquid phase to dynamically refine the mesh, ensuring the smallest element sizes are generated at the evolving interface, during the coalescence process.

After post-processing the simulation data, the in-plane velocity ( $V_{dx}$) and out-of-plane velocity ($V_{dy}$) of the merged droplet are calculated at the moment of detachment from the wall and compared with the corresponding experimental values. The velocities obtained from simulations using the fine and finest meshes are in good agreement with experimental results, as presented in **Table 5.1**. Therefore, the fine mesh configuration is selected for all subsequent simulation studies, where the 20 μm size mesh is generated inside the air domain, and after two levels of refinement, the mesh is generated inside the liquid domain.

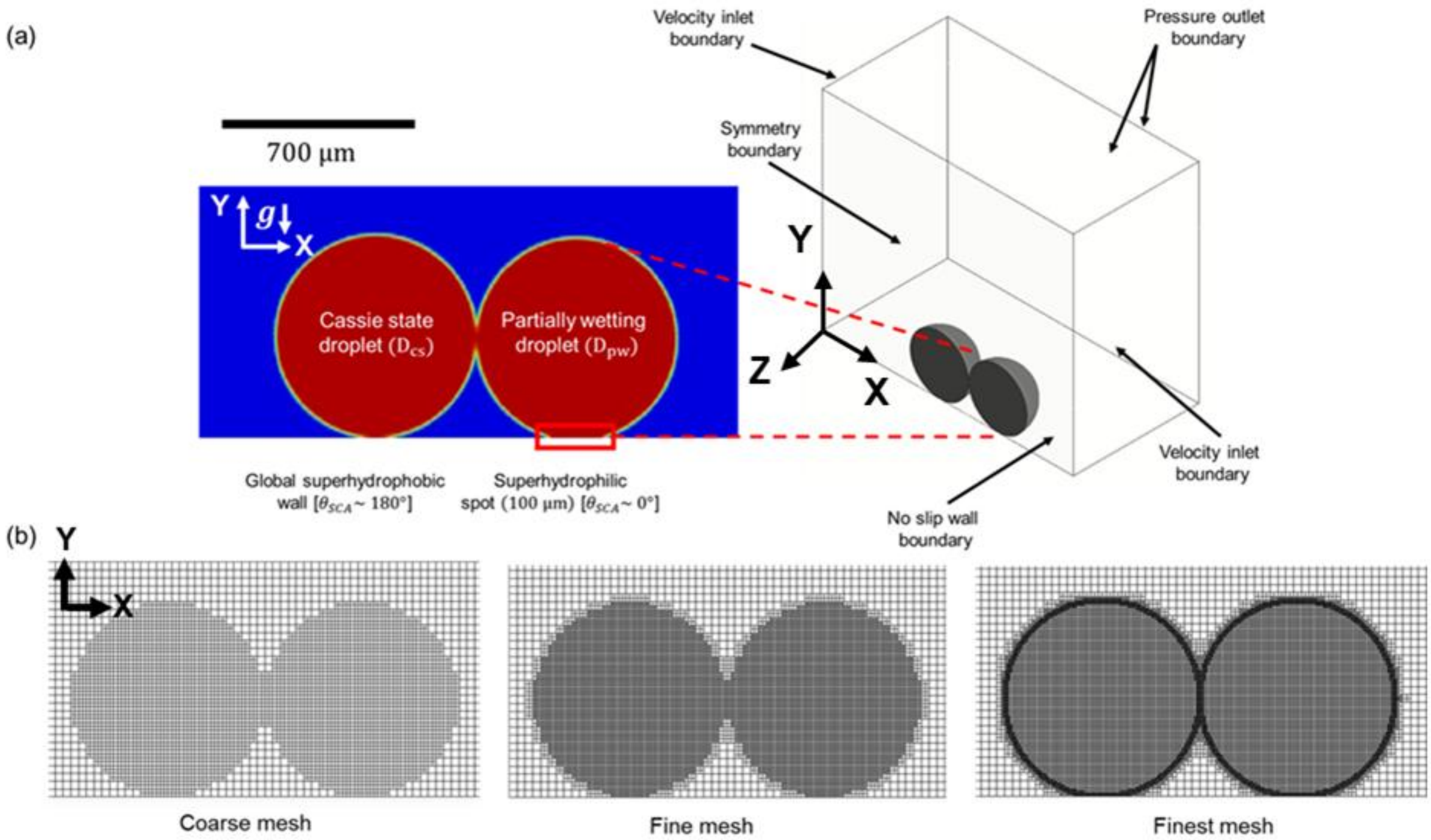

**Figure S.5.** (a) The 3-D computational domain and boundary conditions applied for Cassie state (CS) and partially wetted (PW) droplets coalescence are shown. The inset image highlights the liquid and air phase domains on the symmetry plane. The red region represents the liquid domain, while the blue region indicates the air domain. The superhydrophilic spot is marked with a red rectangular box. (b) Coarse, fine, and finest mesh configurations are shown at the symmetry plane (X-Y plane) of the computational domain.

The morphological evolution during CS–PW droplet coalescence, obtained from numerical simulations with $S^* = 7$, using the fine mesh, is compared with experimental observations, as shown in **Figure S.6**. The numerical simulation closely matches the experimental observations, demonstrating the model's accuracy in capturing the droplet coalescence behavior. Furthermore, there is a strong agreement between the coalescence time predicted by the simulation and that measured in the experiments.

**Table 5.1:** In-plane velocity ($V_{dx}$) and out-of-plane velocity ($V_{dy}$) obtained under different mesh conditions, compared with experimental values.

| Velocity | Coarse mesh | Fine mesh | Finest mesh | Experiment |
|---|---|---|---|---|
| $V_{dx}$ (m/s) | 0.016 | 0.017 | 0.017 | 0.027 |
| $V_{dy}$ (m/s) | 0.069 | 0.100 | 0.114 | 0.107 |

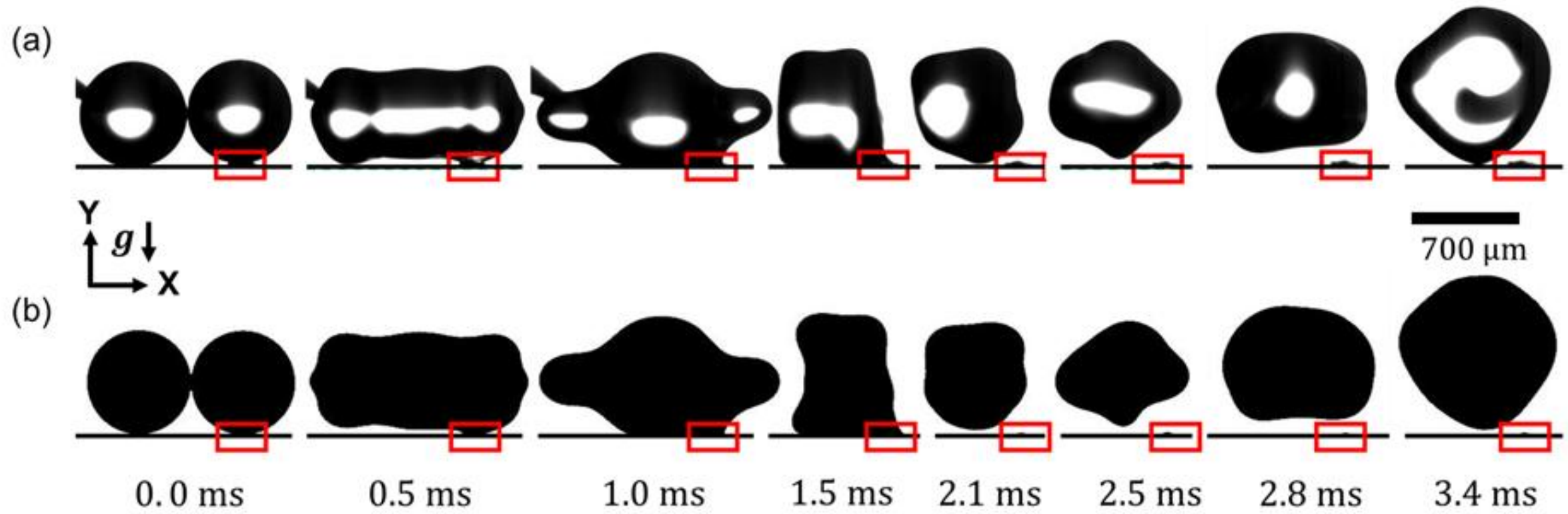

**Figure S.6.** (a) Experimental result and (b) numerical simulation of the coalescence between a 700 µm diameter Cassie droplet (CS) and partial wetting (PW) droplets. The hydrophilic spot ($d_{sh}$), 100 µm in size, is highlighted with a red rectangular box. All observations are made in the X-Y plane (the coordinate system is defined in **Fig. S.5**).

**Subsection S3.2: Flow variables and droplet detachment direction calculation**

With the experimentally validated numerical model, employing the fine mesh strategy and a superhydrophilic spot diameter of $d_{sh} = 100$ µm**,** a series of numerical simulations is performed by varying the sizes of the Cassie state ( $d_{cs}$) and the size of partially wetted ( $d_{pw}$) droplets. The numerical results are post-processed to compute relevant flow variables.

The expressions used for the calculation of these flow variables are provided below.

1) Centre - of - mass velocity

The in-plane centre of mass velocity ($V_{dx}$) at the point of jump is calculated as,

$$V_{dx} = \frac{\int_{\Omega}(\rho\alpha v_x)d\Omega}{\int_{\Omega}(\rho\alpha)d\Omega}$$

Where $d\Omega$ represents the volume of a computational cell within the domain $\Omega$, $\rho$ is the density of the liquid, $v_x$ is the velocity component in the x-direction, and $\alpha$ denotes the volume fraction of liquid in the cell. The integrals are evaluated over the entire computational domain $\Omega$. The out-of-plane velocity ($V_{dy}$) is calculated similarly, replacing $v_x$ with $v_y$.

2) Momentum

The x-momentum of the liquid body is calculated as,

$$p_x = \int_{\Omega}(\rho\alpha v_x)d\Omega$$

The y-momentum ($p_y$) of the liquid body is calculated similarly, replacing $v_x$ with $v_y$.

3) Jumping direction

The jumping direction, calculated by the droplet departure angle ($\theta_d$), is obtained through,

$$\theta_d = tan^{-1}\left(\frac{V_{dy}}{V_{dx}}\right)$$

**Subsection S3.2: Equal–sized droplet coalescence**

The coalescence behavior between same-sized Cassie state (CS) and partially wetted (PW) droplets is explored in this study. Similar to the main text, the numerical results obtained from the coalescence mechanism of CS–PW droplets are characterized using a non-dimensional size parameter $S^*$. Using this parameter, the coalescence dynamics are further analyzed by examining the velocity and detachment angle of the coalesced droplet as functions of $S^*$. Similar to the experimental result, the numerical results presented in **Fig. S.7** reveal a strong dependence of the coalescence behavior on $S^*$.

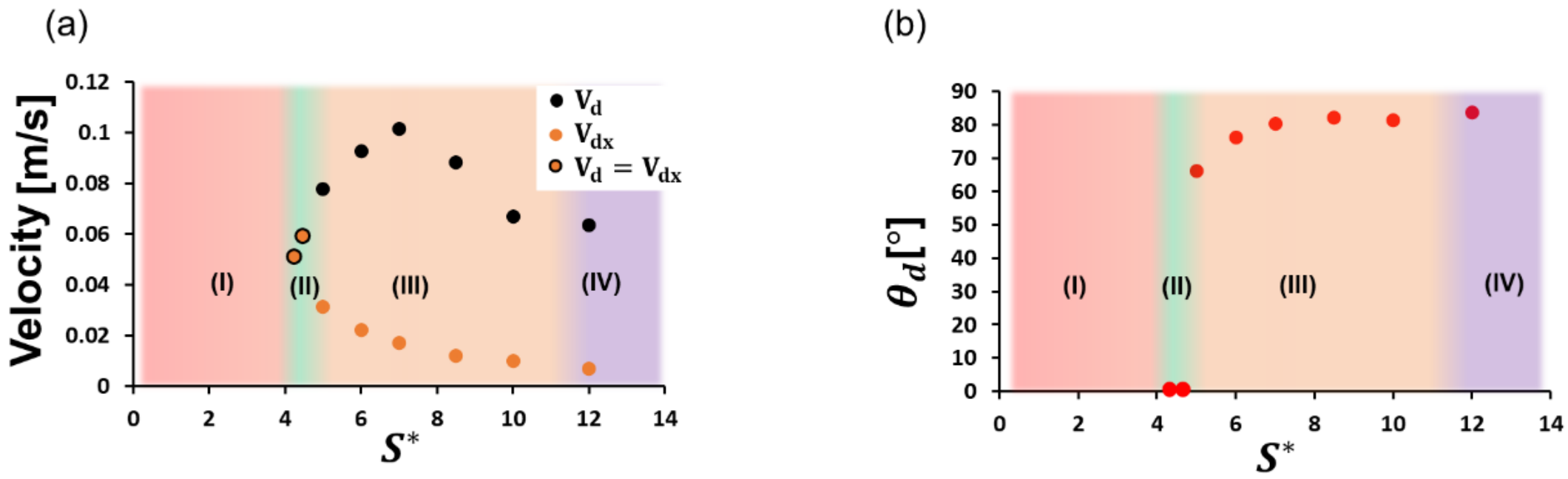

**Figure S.7.** Numerical results for the coalescence of same-size CS and PW droplets: (a) Coalescence-induced droplet total velocity ($V_d$) and in-plane velocity ($V_{dx}$) at the moment of detachment from the superhydrophilic spot as a function of $S^*$. (b) Droplet departure direction

$\theta_d$ as a function of $S^*$. The four distinct coalescence dynamics regimes are marked with I, II, III, and IV. The transition zones between the different regimes are not sharp.

The numerical results reveal three distinct regimes of coalescence dynamics:

- **Pinned Coalescence Regime (Regime I):** The coalesced droplet remains pinned to the superhydrophilic spot without detachment.
- **Lateral Detachment Regime (Regime II):** The coalesced droplet detaches, and complete in-plane motion occurs.
- **Oblique Ejection Regime (Regime III):** The droplet detaches and is ejected at an oblique angle due to a combination of in-plane and out-of-plane momentum.
- **Normal Ejection Regime (Regime IV):** The droplet detaches and ejects nearly vertically from the wall, driven primarily by out-of-plane motion.

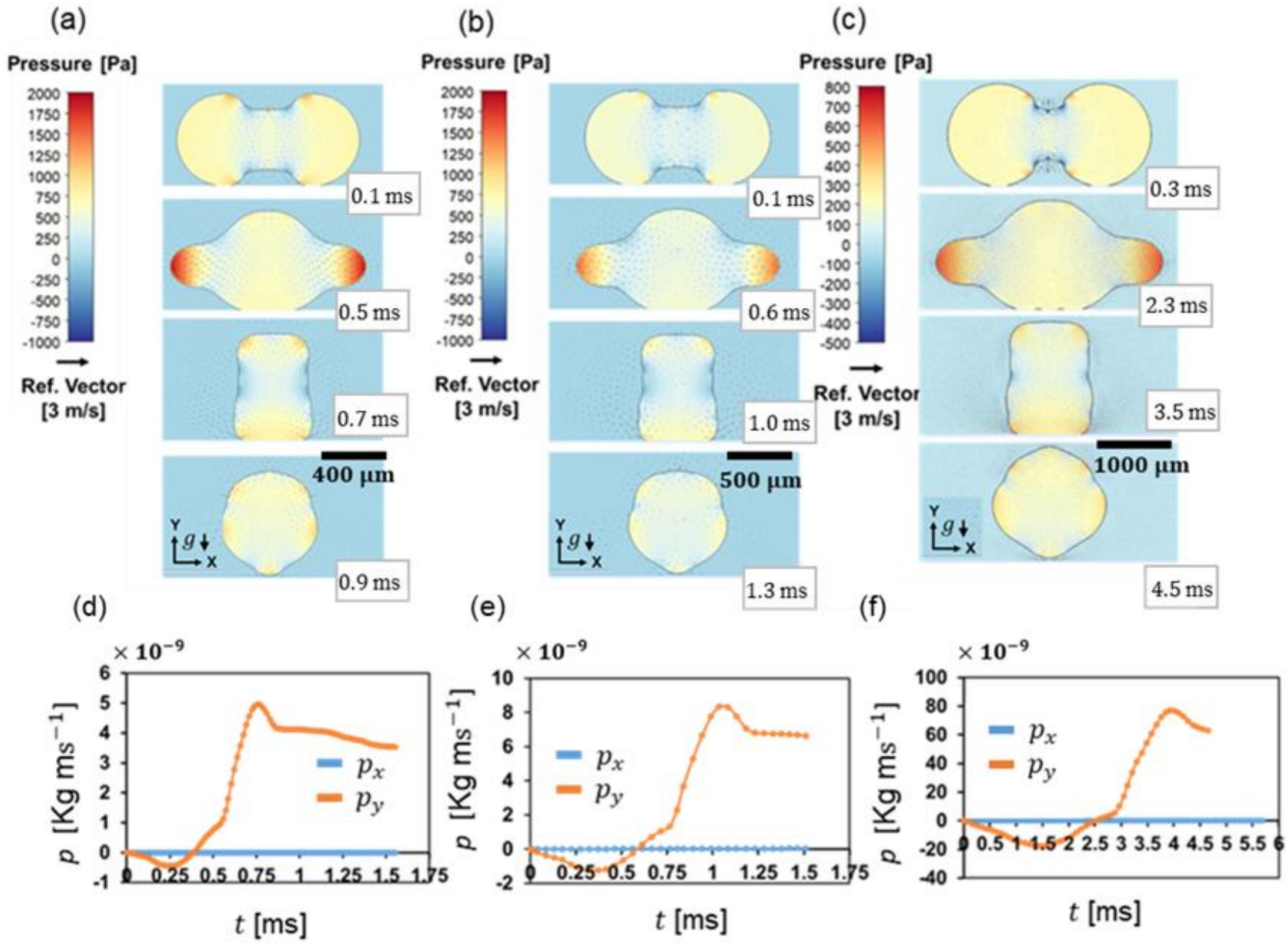

**Figure S.8.** Temporal evolution of velocity and pressure fields at the symmetry plane during equal-sized CS–CS droplet coalescence: (a) 400 μm, (b) 500 μm, and (c) 1200 μm droplets. Corresponding temporal variations of in-plane momentum ($p_x$) and out-of-plane momentum ($p_y$) are shown in (d), (e), and (f), respectively.

Compared to the experimental observations, the pinned coalescence regime (Regime I) was found for $S^*$ values below 3, while in the numerical investigation, this regime was observed for $S^*$ value below 4. While, the lateral detachment regime (Regime II) was found for $S^*$ values between ~ 3 to ~3.5, while in the numerical investigation, this regime was observed for $S^*$ value between ~ 4.1 to ~4.3. Unlike the experiments, where the oblique ejection regime (Regime III) was identified for $S^*$ between ~ 4.5 to ~8, the simulations revealed this regime, with a jumping angle, with $\theta_d < 85°$, for a wider range of $S^*$ between ~ 5 to ~12. Beyond this range, the normal ejection regime (Regime IV) is observed, where $V_{dx} \sim 0$ and $\theta_d \sim 90°$.

To compare the hydrodynamics variation with and without the presence of a pinning spot during equal-size droplets coalescence. Simulations were performed for the coalescence of equal-sized CS-CS droplets, keeping the droplet size the same as in the CS-PW coalescence illustrated in **Fig. 5** of **the** main text. As shown in **Fig. S.8,** for all three cases, symmetrical distribution of pressure and velocity fields develops within the liquid body during coalescence and negligible in-plane momentum ($p_x$) is generated.

**Subsection S3.4: Unequal–sized droplet coalescence**

To examine the impact of size mismatch on coalescence hydrodynamics, additional simulations are conducted using Cassie state (CS) and partially wetted (PW) droplets of unequal radii. From previous observations involving equal-sized CS–PW droplets, it is found that the droplet exhibits more oblique jumping behavior at a non-dimensional size ratio of $S^* \sim 5$. Therefore, to isolate the effect of size mismatch, a fixed value of $S^* = 5$ is maintained, while the diameter of the CS droplet ($d_{cs}$)is varied. The resulting hydrodynamic behavior is analyzed using the non-dimensional diameter ratio, defined as $D^* = \frac{d_{pw}}{d_{cs}}$ . Similar to the experimental observations, the numerical simulations capture both the pinned coalescence regime (Regime I) and the oblique ejection regime (Regime III) and coalescence dynamics at extreme size mismatch $D^* = 0.5$. The pinned regime is observed at $D^* = 1.25$, while the oblique ejection regime occurs at $D^* = 0.83,$ which is consistent with the experimental findings described in the main text. The coalescence dynamics at extreme droplet size mismatch $D^* = 0.5$, reveal complex and highly asymmetric pressure and velocity distribution fields during the coalescence process, which differ significantly from the previously classified regimes.

The hydrodynamics and the associated variation in in-plane momentum ( $p_x$) and out-of-plane momentum ( $p_y$) are compared for each of the coalescence dynamics regimes as illustrated in **Fig. S.9**. In Regime I, as shown in **Figs. S.9(a,d)**. In this case, the CS droplet is smaller than the PW droplet, resulting in the formation of higher pressure in the left lobe of the merged droplet (relative to the larger PW droplet) shortly after coalescence, as shown at 0.6 ms. Afterwards, the coalescing droplet shows similar dynamics as described in the equal-sized CS-PW droplet coalescence for Regime I as shown in **Fig. 5(a)**. As a consequence, the pressure in the left lobe of the merged droplet is higher than the right lobe, reinforcing the momentum imbalance and leading to a continued rise in $p_x$. The in-plane momentum ($p_x$) reaches its peak at ~ 3.5 ms. The pinned droplet state at the peak of $p_x$ is shown at 3.5 ms. The hydrodynamics of the oblique ejection regime (III) are illustrated in **Figs. S.9 (b,e)**, up to the point where the droplet detaches from the wall. The early-stage coalescence dynamics remain similar to the previous case up to 0.8 ms, with the key difference being the asymmetrical pressure distribution between the left and right lobes. In this case, the right lobe exhibits higher pressure than the left, as the PW droplet is smaller than the CS droplet. Afterwards, the coalescing droplet shows similar dynamics as described in the equal-sized CS-PW droplet coalescence for Regime III.

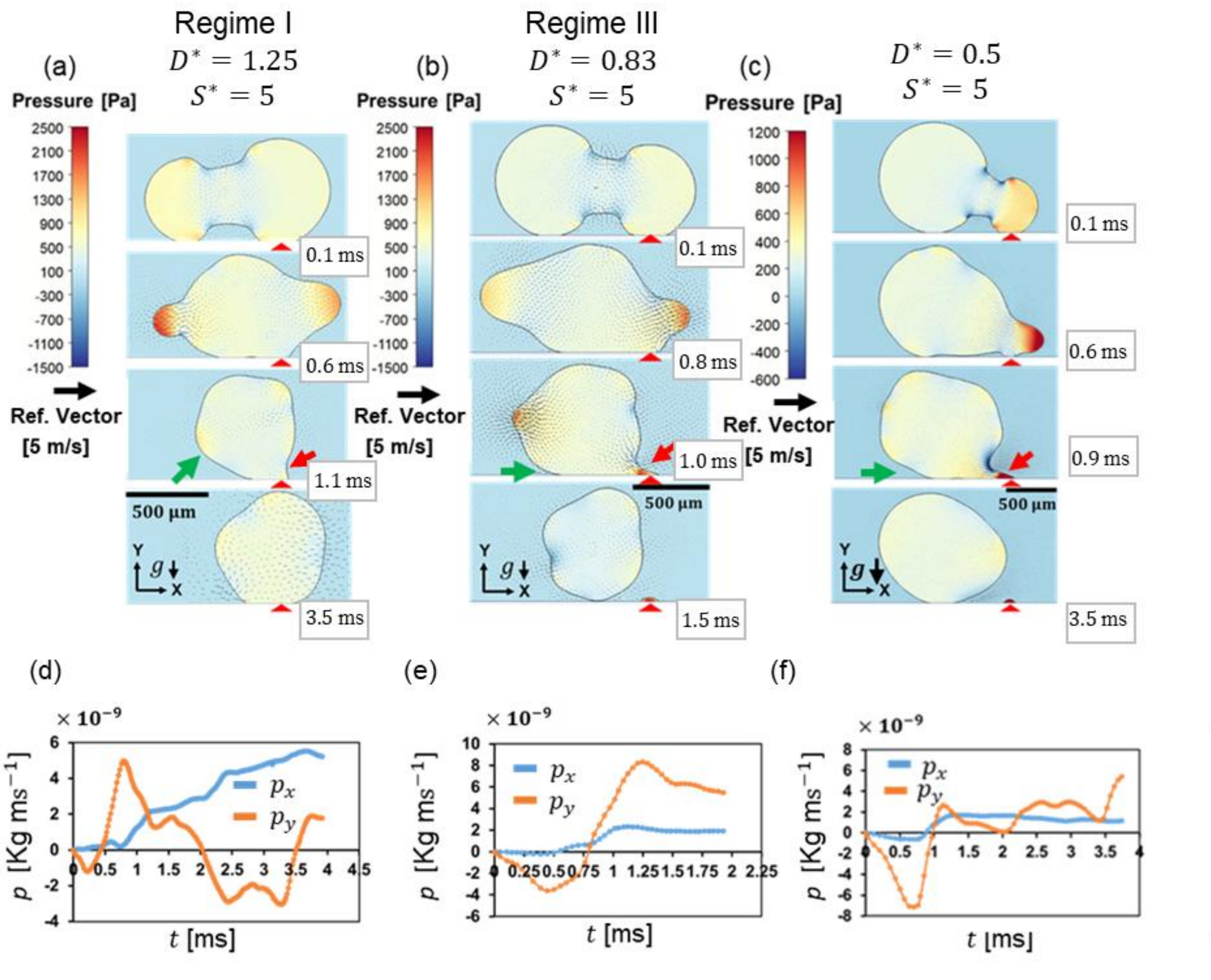

**Figure S.9.** Temporal evolution of velocity and pressure fields at the symmetry plane during coalescence of unequal-sized CS–PW droplets: (a) 400 µm (left) - 500 µm (right) droplets in the pinned coalescence regime (Regime I), (b) 600 µm (left) - 500 µm (right) droplets in the oblique ejection regime (Regime III), and (c) 100 µm (left) - 500 µm (right) droplets. The superhydrophilic spot is marked by a red upward arrow. Corresponding temporal variations of in-plane momentum ($p_x$) and out-of-plane momentum ($p_y$) are shown in (d), (e), and (f), respectively.

It is observed that at 1.0 ms, the curvature of the merged droplet evolves such that a negative pressure region develops near the pinned edge (indicated with a red arrow), while a positive pressure region forms at the opposite edge (indicated with a green arrow). The detached droplet jumps off obliquely from the wall at 1.5 ms. The hydrodynamics of CS–PW droplet coalescence under extreme size asymmetry ($D^* = 0.5$) are illustrated in **Figs. S.9(c,f)**. In this case, the PW droplet is significantly smaller than the CS droplet, which leads to the development of high pressure in the right lobe of the merged droplet, as shown at 0.6 ms. The merged droplet detaches from the pinning spot like that described in the previous regime. However, unlike in Regime III, the merged droplet does not jump off the wall. The state of the merged droplet after a long time following detachment is shown at time 3.5 ms. The overall dynamics are consistent with the experimental observations for these three cases, as described in the main text. The momentum variation plot reveals that both in-plane momentum ($p_x$) and out-of-plane momentum ($p_y$) are generated in the pinned coalescence regime (Regime I), with $p_x$ being larger than $p_y$, as shown in **Fig. S.9(d)**. In the oblique ejection regime (Regime III), the generation of $p_x$ is smaller than $p_y$ as shown in **Fig. S.9(e)**. For the case of extreme size mismatch ( $D^* = 0.5$), the magnitudes of both $p_x$and $p_y$ are comparatively much smaller than those observed in Regimes I and III, as shown in **Fig. S.9(f)**.

To compare the hydrodynamics variation with and without the presence of a pinning spot during unequal-size droplets coalescence. Simulations were performed for the coalescence of unequal-sized CS-CS droplets, keeping the droplet size mismatch ratio the same as in the CS-PW coalescence illustrated in **Fig. S.9**. As shown in **Fig. S.10,** for all three cases, although an asymmetrical distribution of pressure and velocity fields develops within the liquid body during coalescence, no in-plane momentum ($p_x$) is generated. This result is consistent with previous literature, which also reports the absence of $p_x$ generation during the unequal size CS-CS droplets coalescence.[3]

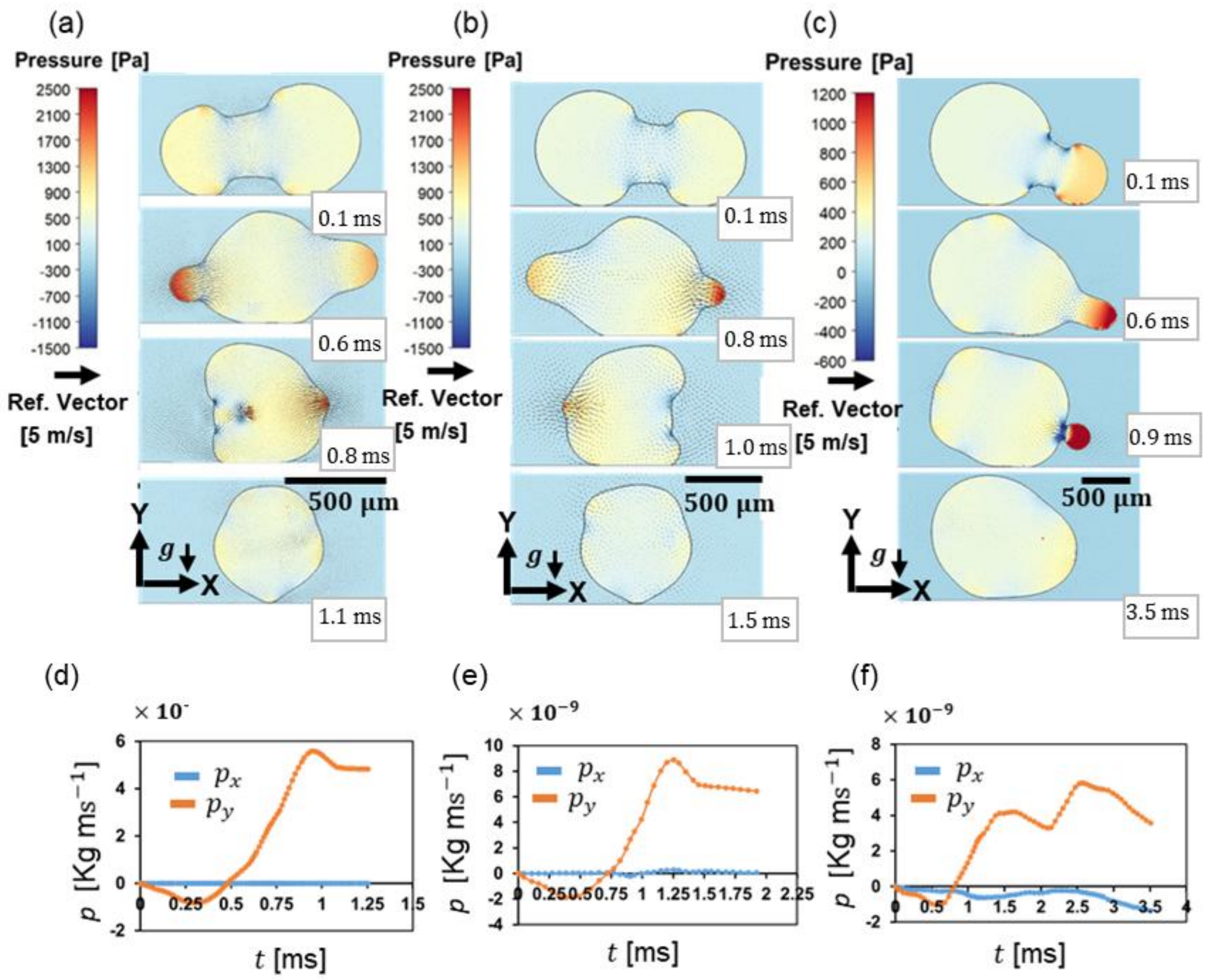

**Figure S.10.** Temporal evolution of velocity and pressure fields at the symmetry plane during coalescence of unequal-sized CS–CS droplets: (a) 400 µm (left) - 500 µm (right) droplets, (b) 600 µm (left) - 500 µm (right) droplets, and (c) 100 µm (left) - 500 µm (right) droplets. Corresponding temporal variations of in-plane momentum ( $p_x$) and out-of-plane momentum ( $p_y$) are shown in (d), (e), and (f), respectively.